\documentclass[utf8]{frontiersFPHY} 

\usepackage{url,hyperref,lineno,microtype,subcaption}
\usepackage[onehalfspacing]{setspace}

\newcommand{\gsim}{\raisebox{-0.13cm}{~\shortstack{$>$ \\[-0.07cm]
			$\sim$}}~}

\def\be{\begin{eqnarray}}
\def\lam{\lambda}
\def\non{\nonumber}
\def\ed{\end{eqnarray}}

\linenumbers

\def\keyFont{\fontsize{8}{11}\helveticabold }
\def\firstAuthorLast{Arhrib {et~al.}} 
\def\Authors{A.~Arhrib\,$^{1}$, R.~Benbrik\,$^{2*}$, H.~Harouiz\,$^{2}$, S.~Moretti\,$^{3}$ and A.~Rouchad\,$^{2}$}
\def\Address{
$^{1}$Facult\'e des Sciences et Techniques, Abdelmalek Essaadi University, B.P. 416, Tangier, Morocco \\
$^{2}$Laboratoire de Physique fondamentale et appliquée – Safi (LPFAS), Facult\'e Polydisciplinaire de Safi, Sidi Bouzid, BP 4162,  Safi, Morocco \\
$^{3}$School  of  Physics  and  Astronomy,  University  of  Southampton, Southampton,  SO17  1BJ,  United  Kingdom}
\def\corrAuthor{Corresponding Author}

\def\corrEmail{r.benbrik@uca.ac.ma}

\begin{document}
\onecolumn
\firstpage{1}

\title[A Guidebook to Hunting Charged Higgs Bosons at the LHC]{A Guidebook to Hunting Charged Higgs Bosons at the LHC} 

\author[\firstAuthorLast]{\Authors} 
\address{} 
\correspondance{} 

\extraAuth{}

\maketitle

\begin{abstract}

\section{}
We perform a comprehensive global analysis in the Minimal Supersymmetric Standard
Model (MSSM) as well as in the 2-Higgs Doublet Model (2HDM) of the production and decay mechanisms 
of charged Higgs bosons $(H^\pm)$ at the Large Hadron Collider (LHC). We start from  accounting for the most recent 
experimental results (SM-like Higgs boson signal strengths and  search limits for new Higgs boson states 
obtained at Run-1 and -2 of the LHC and previous colliders),  from (both direct and indirect)  searches for 
supersymmetric particles as well as from flavor observables (from both $e^+e^-$ factories and hadron 
colliders). We then present precise predictions for $H^\pm$ cross sections and decay rates 
in different reference scenarios of the two aforementioned models in terms of the parameter space 
currently available, specifically, mapped over the customary $(m_{A,H^\pm}, \tan\beta)$ planes.
These include  the $m_{h}^{{\rm mo}d+}$ and hMSSM configurations of the MSSM and  the 2HDM  
Type-I, -II, -X and -Y for which we also enforce theoretical constraints such as vacuum stability, perturbativity and unitarity. 
We also define specific Benchmark Points (BPs)  which are always close to (or coinciding with)  
the best fits of the  theoretical  scenarios to experimental data.
We finally briefly discuss the ensuing phenomenology for the purpose of aiding future searches for 
such charged Higgs boson states.

\tiny
 \keyFont{ \section{Keywords:} Beyond Standard Model, Higgs Physics , Charged Higgs, 2HDM, MSSM, LHC} 
\end{abstract}

\section{Introduction}  

The Higgs boson discovery of 2012~\cite{Aad:2012tfa,Aad:2013wqa,Chatrchyan:2012xdj,Chatrchyan:2013lba}
at the CERN Large Hadron Collider (LHC) has led to
the confirmation of  the Standard Model (SM) as the proper  theory of 
the Electro-Weak (EW) scale. However, there is much evidence that the SM is not appropriate at all scales, rather it  
should be viewed as an effective low-energy realization of a more complete 
and  fundamental theory on setting  beyond the EW regime.  Among the many proposals for the latter, one can list theories with some new symmetry, e.g., 
Supersymmetry (SUSY), or an enlarged particle content (e.g., in the Higgs sector), or both. Following the {aforementioned} discovery, no
new particle has however been seen at the LHC, implying that new
physics at the EW scale should be weakly interacting or that strongly
interacting particles, if present, should lead to signatures involving soft
decay products or in channels with overwhelming  (ir)reducible backgrounds. We shall adopt here the first assumption. 
Many SM {extensions} possess in their spectra additional neutral and/or charged  Higgs states. Amongst these, 
SUSY \cite{Martin:1997ns} is indeed considered  the most appealing one as it addresses several shortcomings of the SM, 
including the problem of the large hierarchy between the EW and Planck scales
{{as well as the dark matter puzzle}}. 
While the search for SUSY was unsuccessful during the first LHC run, the increase in the 
Center-of-Mass (CM) energy of the machine from 8 TeV  to  13 TeV plus the additional luminosity of the second run 
are  improving greatly the sensitivity to the new superparticles which are predicted. While the jury is still out on these, we remind here the reader that SUSY also  requires at least
two Higgs doublets for a successful EW Symmetry Breaking (EWSB) pattern.  For exactly two such fields, yielding the so-called Minimal Supersymmetric Standard Model (MSSM), also having the same gauge group structure of the SM, 
one obtains four  
physical Higgs particles, in addition to the discovered SM-like one with the observed mass of   125 GeV.  In fact, the same Higgs mass spectrum also belongs to a generic 2-Higgs Doublet Model (2HDM), i.e., one not originating from SUSY. In neither case, though, there exists a precise prediction of the typical masses of the new Higgs states, though we know already that the MSSM {{allows for one to be lighter than the 125 GeV state in a very small region of parameter space \cite{Bahl:2018zmf}}}, whereas the 2HDM generally does so over a significantly larger expanse of it  \cite{ATLAS:2014yka,CMS:2014ega}.   Either way, 
the presence  of extra physical Higgs boson states alongside the SM-like one is thus one of the characteristics of Beyond the SM
(BSM) physics, whether within SUSY or otherwise. Hence, looking for these additional states  in various  production and decay channels over
a wide range of kinematic  regimes is an important part of the physics {programme} of the multi-purpose LHC experiments, ATLAS and CMS. Specifically, 
the discovery of a (singly) charged Higgs boson would point to a likely
additional Higgs doublet ({{or a Higgs field with higher
    representation such as triplet}}). Hence, we concentrate on this Higgs state here.
The two Higgs doublet fields pertaining to the MSSM are required to break the EW symmetry and to generate the 
isospin-up and -down type fermions as well as the $W^\pm$ and $Z$ boson masses
\cite{Review0,Carena:2002es,Djouadi:2005gj}. The Higgs spectrum herein is given by the following states: two charged $H^\pm$'s, a CP-odd $A$ and two CP-even Higgses $h$ and $H$, with $m_h<m_H$ (conventionally, wherein $h$ is the SM-like Higgs state). The tree-level  phenomenology of the Higgs sector of the MSSM 
is described entirely by two input parameters, one Higgs mass (that can be taken to be 
that of the CP-odd Higgs state, $m_{A}$) and the ratio $\tan\beta$ of the Vacuum Expectation 
Values (VEVs) of the two Higgs doublet fields.  {{Note that one of the most powerful prediction of SUSY 
is the existence of a light Higgs boson that could be produced at colliders. In the MSSM, at the tree-level, 
the light CP-even $h$ is predicted to be lighter than the $Z$ boson. However, it is well known that loop effects could raise the $h$ mass upper bound to 135 GeV for a large soft breaking trilinear parameter,  $A_t$, and/or a heavy scalar top \cite{CR-1loop,CR-eff,CR-2loop,Degrassi:2002fi}. }}
After the Higgs boson discovery at the LHC, MSSM benchmark scenarios have been refined to match
the experimental data and to reveal characteristic features of certain regions of the parameter
space \cite{Carena:2005ek, Carena:2013ytb, Carena:2002qg}. 
Of the many MSSM frameworks presented in literature, we consider in this
work the so-called $m^{{\rm mod}+}_{h}$ \cite{Carena:2013ytb} and hMSSM \cite{Djouadi:2013uqa,Bagnaschi:2015hka} ones,
 which will be described in the coming section. 
As for the 2HDM, one ought to specify the Yukawa sector, in order to proceed to study {phenomenologically} its manifestations. While SUSY enforces this in the form of a so-called Type-II, this is only one of four Ultra-Violet (UV) complete {realizations} of a generic 2HDM, the others been termed Type-I, -X and -Y. The difference between these four scenarios is the way the fermionic masses are generated. We define as Type-I the model where only one doublet couples to all fermions, Type-II is the scenario where one doublet couples to up-type quarks and the other to down-type quarks and leptons, the Type-X is the model where one doublet couples to all quarks and the other to all leptons while a Type-Y  is built such that one doublet  couples to up-type quarks and to leptons and the other to down-type quarks. In all such cases, the number of free parameters at tree-level is seven to start with, hence it becomes more cumbersome than in SUSY to map experimental results onto theoretical constraints. Yet, in virtue of the fact that a 2HDM is the simplest  {realization} of a BSM scenario based solely on doublet Higgs fields, its study is vigorously being pursued experimentally.

So  far,  the non-observation  of  any Higgs signal events in direct searches above and beyond those of the SM-like Higgs state 
constrains the parameter space of the underlying physics model. Specifically, in the case of the $H^\pm$ boson, wherein the relevant phenomenological parameters are $m_{H^\pm}$ and $\tan\beta$ in whatever scenario, one can pursue the study of its  production and decay modes in a model independent way, which results can a posteriori  be translated to exclude the relevant parameter space in a given scenario (whether it be the MSSM, 2HDM or something else). This recasting is conveniently done on the  $(m_{A}, \tan\beta)$ and  $(m_{H^\pm}, \tan\beta)$  planes for the MSSM and 2HDM, respectively, so that we will map our findings in the same way.

At hadron colliders, there exists many production  modes for charged Higgs bosons which are rather similar in the MSSM and 2HDM. For a light charged Higgs, i.e., with mass $m_{H^\pm}+m_b<m_t$, its production comes mainly from top decay. At the LHC, the production of top quark pairs proceeds via Quantum Chromo-Dynamics (QCD) 
interactions and,   when kinematically allowed, one top could decay into a charged Higgs state and a bottom quark in a competition with the SM decay into a $W^\pm$  boson and again a bottom quark. Therefore, the complete $H^\pm$ production mechanism $q\bar q,gg\to t\bar{t}\to t\bar{b}H^-$ provides the main source of light charged Higgs bosons at the LHC and offers a much more copious  signature than  any  other form of direct production. After crossing the top-bottom threshold, i.e., when $m_{H^\pm}+m_b>m_t$, a
charged Higgs (pseudo)scalar can be produced through the process $gb\to t H^-$  \cite{Barger:1993th,Akeroyd:2016ymd}. In fact, these two mechanisms can be simultaneously captured via the process $gg\to t\overline{b}H^-$  \cite{Borzumati:1999th,Guchait:2001pi}, which again makes it clear that one  should expect large $H^\pm$ cross sections induced by QCD interactions {also} in the heavy $H^\pm$ 
mass range\footnote{For a complete review on charged Higgs production modes, see \cite{Akeroyd:2016ymd}.}.

In the MSSM, and also in a variety of 2HDM Types, light charged Higgs bosons would decay almost 
exclusively into a (hadronic or leptonic)  $\tau$
lepton and its associated neutrino for $\tan\beta \gsim 1$. When the top-bottom channel 
is kinematically open, then $H^+\to t\bar b$ would compete with $H^\pm \to hW^\pm, HW^\pm, AW^\pm$ decays as well as various SUSY channels in the MSSM.
In the latter, $H^+\to t\bar b\to b\bar b W^+$ is the dominant channel and the bosonic decays $H^\pm\to hW^\pm, HW^\pm, AW^\pm$ (also yielding 
$b\bar b W^+$ final states) are subleading.
In the 2HDM, if none of these bosonic decays is open, then  $H^+\to t\bar b$ is the dominant  mode. 
At the LHC Run-1,  lighter charged Higgs bosons were probed in the decay channels $\tau\nu$
\cite{Aad:2014kga,Khachatryan:2015qxa}, $cs$ \cite{Aad:2013hla,Khachatryan:2015uua} and also $cb$ \cite{CMS:2016qoa}. No excess was observed and model independent   limits are set  on the following product of Branching Ratios (BRs): BR$(t\to H^+ b)~\times$ BR$(H^+\to \tau \nu)$. 
At Run-2, mainly the decay modes $\tau\nu$  \cite{Aaboud:2016dig, CMS:2016szv} and $tb$ \cite{ATLAS:2016qiq} are explored in the mass
range $m_{H^\pm}=200$ GeV  to 1000 GeV, in the latter mode  using multi-jet final states with one electron or muon from the top quark decay. No significant excess above the background-only  hypothesis has been observed and upper limits are set on the $pp\to tbH^\pm$ 
production cross section times BR$(H^{\pm}\to tb)$. Several interpretations  of these limits have eventually been given in benchmark scenarios of the MSSM, including those mentioned above.  
Note that  current ATLAS and CMS bounds are significantly weakened in the
2HDM once the exotic decay channels into a lighter neutral Higgs, e.g., $H^\pm \to h W^\pm $ or $H^\pm\to A W^\pm $, are open.  
This scenario could also happen in the MSSM if one of the SUSY decay channels of charged Higgs bosons 
are open (such as into chargino-neutralino pairs). 
In the 2HDM,  the  possibility of producing a light charged Higgs boson from top decay with a subsequent step
$H^\pm \to h W^\pm$ or $H^\pm \to A W^\pm $
was studied in \cite{Arhrib:2016wpw} and it was shown that it can lead to 
sizable cross sections at low $\tan\beta$. 
{{We stress here that there exist several recent analyses dedicated to 2HDM phenomenology 
\cite{Akeroyd:2016ymd,Arbey:2017gmh,Bernon:2015qea} that we consulted. However, unlike Ref. \cite{Akeroyd:2016ymd}
Refs. \cite{Arbey:2017gmh,Bernon:2015qea} only concentrates on neutral Higgs 
phenomenology and discuss the charged Higgs contribution only to flavor physics observables without singling out the relevant 
charged Higgs production and decay channels at the LHC, which is indeed one of the aims of this analysis. }}

In this paper, we analyze the allowed  $\sigma(pp\to t\bar b H^+ + {\rm c.c.})\times{\rm  BR}(H^\pm\to {\rm anything})$ 
rates by taking into account both
theoretical and experiments constraints on the underlying BSM model, the latter including the latest ATLAS and CMS
results for SM-Higgs ($h$) and  other Higgs ($H, A, H^+$) searches with the full set of 36.5 fb$^{-1}$ data 
collected  in the second LHC phase. We will then interpret these results under the proposed
scenarios to quantify the magnitude of the available parameter space to be covered by future LHC analyses. In doing so, we will extract several 
Benchmarks Points (BPs) that could lead to detectable signals, all of which are consistent with the best fit regions in both the MSSM and  2HDM.

The paper is organized as follows. In the second section we review the MSSM  and introduce the benchmark scenarios that we will discuss.
The 2HDM, with its parameterizations and  Yukawa textures, is described in the third section. 
The fourth section is devoted to a discussion of the  theoretical and experimental constraints used in our study. 
Results and discussions for the  MSSM and 2HDM are presented in the fifth section and we finish with our conclusions.

\section{The MSSM}
\label{sec:modelMSSM}
In the MSSM, due to the holomorphy of the superpotential, one introduce 
two Higgs doublets $\Phi_{1,2}$ in order to give masses to up-type quarks as well as down-type quarks and  leptons, respectively.
Both Higgs fields acquire VEVs, denoted by $v_{1,2}$. After EWSB takes place, 
the spectrum of the model contains the aforementioned Higgs states: $h,H,A$ and $H^\pm$.
The MSSM Higgs sector is parameterized at tree-level by $\tan\beta=v_2/v_1$ and, e.g., the CP-odd mass $m_A$.
One of the interesting features of the MSSM is the prediction, at the tree-level, of a  light CP-even Higgs $h$ with a mass $m_h\leq m_Z$. However, such tree-level prediction is strongly modified by radiative corrections at one- and two-loop 
level \cite{CR-1loop, CR-eff,CR-2loop}. It has been shown in \cite{Degrassi:2002fi}, on the one hand, that the loop effects can 
make the light CP-even mass $m_h$  reach a value of 135 GeV  and, on the other hand, that the theoretical uncertainties due 
to  unknown high order effects can be of the order of 3 GeV.
In fact, these large loop effects are welcome in order to shift the light CP-even Higgs mass 
to the measured experimental  value $m_h\approx 125$ GeV. 
Note also that the loop effects will modify not only the tree-level Higgs mass relations  
but also the Higgs self-couplings and the  Higgs coupling to SUSY particles.
Therefore, beside the tree-level parameters $\tan\beta$ and $m_A$, 
the top quark mass and the associated squark masses and their soft SUSY breaking parameters
enter through radiative corrections \cite{CR-1loop, CR-eff, CR-2loop,Heinemeyer:1998jw,Heinemeyer:1998kz}. 
In fact, when trying to  push the light CP-even mass from $m_Z$ to 125 GeV through loop effects, one needs to introduce a  large SUSY scale with large soft trilinear parameter $A_t$. Such a  large SUSY scale puts  automatically the SUSY  
spectrum at the TeV scale,  which is  consistent with  negative searches for SUSY particles at the LHC.

To compute the masses and  couplings of Higgs bosons in a given point of the MSSM parameter space we use the  code
FeynHiggs \cite{Heinemeyer:1998yj,Hahn:2009zz} for the $m_h^{mod+}$ scenario and the  program HDECAY for the  hMSSM case \cite{Djouadi:1997yw}.
Both codes include the full one-loop and a large part of the dominant two-loop  corrections  to  the
neutral Higgs masses.  Since the theoretical uncertainty on the Higgs mass calculation in the 
FeynHiggs code has been estimated to be of the order of 3 GeV, we consider as phenomenologically acceptable
the  points  in  the  MSSM  parameter  space  where FeynHiggs predicts  the  existence  of  a  scalar state with
mass between 122.5 GeV and 128.5 GeV and with approximately SM-like couplings to gauge bosons and fermions. 
In addition to the tree-level scalar potential parameters, $\tan\beta$ and $m_A$, 
when taking into account high order corrections, the MSSM parameters most relevant to the prediction of the masses and production cross sections of the Higgs bosons are:  the soft SUSY-breaking masses for the stop and sbottom squarks, which, for simplicity, we assume all equal to a common mass parameter $M_{\rm SUSY}$, the soft
SUSY-breaking gluino mass $m_{\tilde {g}}$, the soft SUSY-breaking Higgs-squark-squark couplings $A_t$ and $A_b$, the  superpotential Higgs(ino)-mass parameter $\mu$ and the left-right
mixing terms in the stop and sbottom mass matrices  {{(divided by $m_t$ and $m_b$)}}  
\begin{eqnarray}
X_t = A_t - \mu \cot\beta,\qquad \qquad X_b = A_b - \mu\tan\beta,
\end{eqnarray}
respectively. Since the (approximate) two-loop calculation of the Higgs masses implemented in
FeynHiggs and the Next-to-Leading Order (NLO) calculation of QCD corrections to  
the production cross section implemented in SusHi~\cite{Harlander:2012pb,Harlander:2016hcx}
employ the same {renormalization}
(on-shell)  scheme, the input values of the soft SUSY-breaking parameters can be passed seamlessly from the Higgs mass  to the cross section calculations.

In the light of the latest LHC data on the discovered Higgs-like boson,  and 
given the fact that the MSSM contains many independent  parameters which makes it a fastidious task to perform a full scan,
there have been many  studies which lead to several benchmarks that could fit the observed Higgs boson as well as be 
tested at the future LHC with higher luminosity \cite{Carena:2013ytb,Djouadi:2013uqa} .
As intimated already, in this study, we will concentrate on two of these benchmark scenarios: the $m_h^{{\rm mod}+}$ and hMSSM ones, which we will describe hereafter.

\subsection*{The $m^{{\rm mod}+}_{h}$ scenario}
The $m^{{\rm mod}+}_{h}$ scenario is a modification of the time-honoured
$m^{\rm max}_{h}$ scenario (also called maximal mixing scenario), which was originally defined  to give
conservative exclusion bounds on $\tan\beta$ in the context of Higgs boson searches at LEP \cite{Heinemeyer:1999zf}. The 
$m^{\rm max}_{h}$ scenario was introduced in order to maximize the value of $m_h$ by incorporating 
large radiative correction effects for a large $m_A>>m_Z$ mass, fixed value of $\tan\beta>8$ and large SUSY scale of the order 1 TeV.
However, this scenario predicts  $m_h$ to be  much higher  than the observed Higgs boson mass,  due to the 
large mixing in the scalar top sector.

Hence, the maximal mixing scenario has been  modified, by reducing the amount of scalar top mixing, 
 such that the mass of the lightest Higgs state,  $m_h$,  is compatible with the mass of the
observed Higgs  boson within $\pm 3$ GeV in a large fraction of the considered
parameter space.  In fact, modifications of the $m^{\rm max}_{h}$ scenario can be done in two ways depending on the sign of 
$ (A_t -\mu \cot\beta)/M_{\rm SUSY}$,  leading to an $m^{{\rm mod}-}_{h}$ and $m^{{\rm mod}+}_{h}$ 
\cite{Carena:2013ytb}. It has been demonstrated in \cite{Carena:2013ytb}
that when $m^{{\rm mod}+}_{h}$  is confronted with  LHC data, there is a substantial region in the plane 
$(m_A,\tan\beta)$ with $\tan\beta>7$ for which the light CP-even Higgs mass is in  a good agreement with the measured one  at the LHC, hence our choice of this scenario.

The SUSY input parameters in this scenario are fixed as\footnote{{{Notice that this $m^{{\rm mod}+}_{h}$  configuration is compliant with the theoretical and experimental constraints discussed below, including Dark Matter (DM) ones. So is the case for the hMSSM configuration below.}}.}
\begin{eqnarray}
&M_{Q_3}=M_{U_3}=M_{D_3}=1.5~\text{TeV},\quad
M_{L_3}=M_{E_3}=2~\text{TeV}, \nonumber\\[2mm]
&  \mu=200~\text{GeV},\,\quad
M_1=100~\text{GeV},\quad M_2=200~\text{GeV},
\quad m_{\tilde{g}}=1.5~\text{TeV}, \nonumber\\[2mm]
&X_t=2M_{\rm SUSY}=1~{\rm TeV},\quad A_b=A_\tau=A_t,
\label{eq:firstscenario}
\end{eqnarray}
where $M_{\rm SUSY}$ is the aforementioned SUSY mass scale.

\subsection*{The hMSSM scenario}
In the previous scenario, one need to input $\tan\beta$, $m_A$ and also the other SUSY parameters
to get the Higgs and SUSY (mass and coupling) spectrum. Taking into account the theoretical uncertainty of the order 3 GeV 
on the Higgs mass, which could originate from unknown high order loop effects, a light CP-even 
Higgs  boson with a mass  in the range $[122,128]$ GeV would be an MSSM candidate for the observed Higgs-like particle.
However, plenty of points on the $(m_A,\tan\beta)$ plane would correspond to  one configuration of  $m_h$ mass.
To avoid this situation, the hMSSM benchmark was introduced \cite{Djouadi:2013uqa}. In this scenario, the light CP-even Higgs state 
is enforced to be approximately 125 GeV while setting the SUSY mass scale $M_{\rm SUSY}$ to be rather high (i.e., $>1$ TeV) in order to 
explain the non-observation of any SUSY particle at colliders.
The hMSSM setup thus describes the MSSM Higgs sector in terms
of just $m_A$ and $\tan\beta$, exactly like for tree-level predictions, given the experimental knowledge of $m_Z$ and $m_h$. 
In this scenario, therefore, the dominant radiative corrections would be fixed by the measured experimental value of $m_h$ 
which in turn fixes the SUSY scale  \cite{Djouadi:2013uqa} .
It defines a largely model-independent scenario, because the
predictions of the properties of the MSSM Higgs bosons do not depend 
on  the  details  of  the  SUSY  sector, somewhat unlike the previous case, {wherein} 
squark masses are fine-tuned to obtain $m_h\approx125$ GeV.  

The SUSY input parameters in this scenario are similar to the 
previous one, Eq.~(\ref{eq:firstscenario}), except that we take $X_t=2M_{\rm SUSY}=2$ TeV.\\

\subsection*{Setup}
Both scenarios introduced above are characterized by relatively large values of the ratio $X_t /M_{\rm SUSY}$, 
ensuring that the MSSM mass of the SM-like Higgs state falls 
within the required range without the need for an extremely heavy stop.   In addition, the masses of the gluino and   first two 
generation squarks are set to 1.5 TeV, large enough to evade the current ATLAS and CMS  limits stemming from SUSY searches 
\cite{ATLAS:2013tma, Aad:2013wta,Chatrchyan:2013mys,Chatrchyan:2013fea,Chatrchyan:2014lfa}. 
We vary the parameters $\tan\beta$ and $m_A$ within the following 
ranges:
\begin{equation}
0.5 \le \tan\beta \le 15, \quad\quad 90~{\rm GeV} \le m_A \le 1~{\rm  TeV}.
\end{equation}
The soft trilinear term $A_t$ is set to be equal to $A_b$. Due to the smallness of the light quarks masses, 
the left-right mixing of the first two generation squarks is neglected. 
The gaugino mass parameters $M_1$, $M_2$ and the soft SUSY-breaking gluino mass $m_{\tilde {g}}$
are all related through {Renormalization} Group Equation (RGE) running to some common high scale $m_{1/2}$ soft term 
which yields the relations $m_{\tilde {g}} \approx  3.5M_2$ and $M_1\approx 0.5M_2$. 
In our analysis, we assume Grand Unified Theory (GUT) relations only between $M_1$ and $M_2$ while $M_2$ and $m_{\tilde {g}}$ are taken independent from each other. 
Finally,  the soft SUSY-breaking parameters in the slepton sector have a very small impact on the predictions for the Higgs masses and production cross sections, therefore we do not report on them here.
\section{The 2HDM}
\label{sec:model2HDM}
In this section,   we define  the scalar potential and the Yukawa sector of 
the general 2HDM. The most general scalar potential which is 
$SU(2)_L\otimes U(1)_Y$ invariant is given by \cite{Gunion:2002zf,Branco:1999fs}
\be
V(\Phi_1,\Phi_2)
&=& m^2_1 \Phi^{\dagger}_1\Phi_1+m^2_2 \Phi^{\dagger}_2\Phi_2 -(m^2_{12}
\Phi^{\dagger}_1\Phi_2+{\rm h.c}) +\frac{1}{2} \lam_1 (\Phi^{\dagger}_1\Phi_1)^2 
\nonumber \\ &+& \frac{1}{2} \lam_2
(\Phi^{\dagger}_2\Phi_2)^2 +
\lam_3 (\Phi^{\dagger}_1\Phi_1)(\Phi^{\dagger}_2\Phi_2) + \lam_4
(\Phi^{\dagger}_1\Phi_2)(\Phi^{\dagger}_1\Phi_2)  \non \\
&+&  \left[\frac{\lam_5}{2}(\Phi^{\dagger}_1\Phi_2)^2 +{\rm H.c.} \right].
\label{higgspot}
\ed
The complex (pseudo)scalar doublets $\Phi_i$ ($i=1,2$) can be parameterized as
\begin{align}
\Phi_i(x) = \begin{pmatrix}
\phi_i^+(x) \\ 
\frac{1}{\sqrt{2}}\left[v_1+\rho_1(x)+i \eta_1(x)\right]
\end{pmatrix}, 
\end{align}
with $v_{1,2}\geq 0$ being the VEVs satisfying $v=\sqrt{v_1^2+v_2^2}$, 
with $v=246.22$~GeV.
Hermiticity of the potential forces $\lambda_{1,2,3,4}$ 
to be real while $\lambda_{5}$ and $m^2_{12}$ can be complex.
In this work we choose to work in a CP-conserving
potential where both VEVs are real and so are  
also $\lambda_{5}$ and $m^2_{12}$.

After EWSB, three of the eight degrees 
of freedom in the Higgs sector of the 2HDM are eaten by the Goldstone
bosons ($G^\pm$ and $G$) to give masses to the longitudinal gauge bosons ($W^\pm$ and $Z$). 
The remaining five degrees of freedom become the aforementioned 
physical Higgs bosons. After using the {minimization} conditions for the potential
together with the $W^\pm$ boson mass requirement, we end up with seven independent parameters
which will be taken as
\begin{equation}
m_{h}\,, m_{H}\,, m_{A}\,, m_{H^\pm}\,, \alpha\,, \tan\beta\,,  m^2_{12}, 
\label{parameters} 
\end{equation}
where, as usual, $\tan\beta \equiv v_2/v_1$ and $\beta$ is also the angle that {diagonalizes}
the mass matrices of both the CP-odd  and  charged Higgs sector while the angle
$\alpha$ does so in the CP-even Higgs sector. 

{The most commonly used versions of a CP-conserving 2HDM are the ones that satisfy a discrete $Z_2$ {symmetry} 
$\Phi_i \to (-1)^{i+1} \Phi_i$ ($i=1,2$),} that, when extended to the Yukawa sector, guarantees the {absence} of Flavor Changing Neutral Currents (FCNCs). Such a symmetry
would also require $m^2_{12} = 0$, unless we tolerate a soft violation of this by the dimension two term $m^2_{12}$ (as we do here). 
The Yukawa Lagrangian can then be written as
\begin{align}
- \mathcal L_Y=
\bar Q_L( Y_1^d\Phi_1+Y_2^d\Phi_2) d_R 
+\bar Q_L({Y}^u_1 \tilde\Phi_1+{Y}^u_2 \tilde\Phi_2)u_R +
\bar L_L( Y_1^l\Phi_1+Y_2^l\Phi_2) l_R
+\text{H.c.},
\end{align}
where $Q^T_L = (u_L, d_L)$ and $L^T_L=(l_L, l_L)$ are 
the left-handed quark doublet and lepton doublet, respectively, the $Y^f_k$'s 
($k=1,2$ and $f= u,d,l$) denote the $3\times 3$  Yukawa matrices and 
$\tilde\Phi_k = i\sigma_2 \Phi^*_k$ ($k=1,2$). 
The mass matrices of the quarks and leptons are 
a linear combination of  $Y_{1}^f$ 
and $Y_2^f$,  $Y_{1,2}^{d,l}$ and  $Y_{1,2}^u$. Since they   
cannot be {diagonalized} simultaneously in general, 
neutral Higgs Yukawa couplings with flavor violation appear 
at tree-level and contribute significantly to FCNC processes such as $\Delta M_{K, B, D}$ as well as 
$B_{d,s} \to \mu^+ \mu^-$ mediated by neutral Higgs exchanges.
To avoid having those large FCNC processes, one known solution is to extend the  
$Z_2$ symmetry  to the Yukawa sector. When doing so, we ended up with the already discussed  
four possibilities regarding the Higgs bosons couplings to fermions \cite{Branco:1999fs}.

After EWSB, the Yukawa Lagrangian can be expressed in the mass eigenstate basis as follows
\cite{GomezBock:2005hc,Arhrib:2017yby}:
\begin{eqnarray}
{\mathcal L}_Y &=& -\sum_{f=u,d,\ell} \frac{m_f}{v} \left(\kappa_f^h \bar f fh + \kappa_f^H \bar f fH - i \kappa_f^A \bar f \gamma_5 f A \right) \nonumber \\
&& - \Big(\frac{\sqrt 2 V_{ud}}{v} \bar u \left (m_u \kappa_u^A P_L +  m_d\kappa_d^A P_R \right )d H^+ 
+\text{H.c.}\Big).\label{Eq:Yukawa}
\end{eqnarray}

{{We give in Tab. \ref{Yukawa} the values of the Yukawa couplings $\kappa_f^\phi$ ($\phi=h,H,A$)}}, {{i.e., the Higgs boson interactions normalized to the SM vertices introduced in Ref.  \cite{LHCHiggsCrossSectionWorkingGroup:2012nn}}}, in the four 2HDM Types.
\begin{table}
	\begin{center}
		{\renewcommand{\arraystretch}{1.2}
			{\setlength{\tabcolsep}{.3cm}
		\begin{tabular}{||l|l|l|l|l|l|l|l|l|l||}
			\hline \hline
			& $\kappa_u^h$ & $\kappa_d^h$ & $\kappa_l^h$ & $\kappa_u^H$ & $\kappa_d^H$ & $\kappa_l^H$ & $\kappa_u^A$ & $\kappa_d^A$ & $\kappa_l^A$ \\ \hline
			Type-I & $c_\alpha/s_\beta$ & $c_\alpha/s_\beta$& $c_\alpha/s_\beta$ & $s_\alpha/s_\beta$ & $s_\alpha/s_\beta$ & $s_\alpha/s_\beta$ & $c_\beta/s_\beta$ & 
			$-c_\beta/s_\beta$ & $-c_\beta/s_\beta$ \\ \hline
			Type-II & $c_\alpha/s_\beta$ & $-s_\alpha/c_\beta$& $-s_\alpha/c_\beta$ & $s_\alpha/s_\beta$ & $c_\alpha/c_\beta$ & $c_\alpha/c_\beta$ & $c_\beta/s_\beta$ & 
			$s_\beta/c_\beta$ & $s_\beta/c_\beta$ \\ \hline 
			Type-X & $c_\alpha/s_\beta$ & $c_\alpha/s_\beta$& $-s_\alpha/c_\beta$ & $s_\alpha/s_\beta$ & $s_\alpha/s_\beta$ & $c_\alpha/c_\beta$ & $c_\beta/s_\beta$ & 
			$-c_\beta/s_\beta$ & $s_\beta/c_\beta$ \\ \hline
			Type-Y & $c_\alpha/s_\beta$ & $-s_\alpha/c_\beta$& $c_\alpha/s_\beta$ & $s_\alpha/s_\beta$ & $c_\alpha/c_\beta$ & $s_\alpha/s_\beta$ & $c_\beta/s_\beta$ & 
			$s_\beta/c_\beta$ & $-c_\beta/s_\beta$ \\ \hline \hline
		\end{tabular}}}
	\end{center}
	\caption{{{Yukawa couplings in terms of the standard $\kappa$ coefficients, in turn expressed as function of the angles $\alpha$ and $\beta$,  in the four  2HDM Types. Here, the shorthand notation $c_x\equiv\cos x$ and $s_x\equiv\sin x$ is used}}.}
	\label{Yukawa}
\end{table} 
The couplings of $h$ and $H$ to gauge bosons $W^\pm,Z$ are proportional to 
$\sin(\beta-\alpha)$ and  $\cos(\beta-\alpha)$, respectively.
Since these are gauge couplings, they are the same for all Yukawa types.
As we are considering the scenario where the lightest neutral Higgs state is the 125 GeV scalar,
the SM-like Higgs boson $h$ is recovered when $\cos(\beta-\alpha)\approx 0$.
As one can see from Tab. \ref{Yukawa}, for all 2HDM Types,  
this is also the limit where the Yukawa couplings of the discovered Higgs 
boson  become SM-like. The limit $\cos(\beta-\alpha)\approx 0$ seems to be favored by LHC data, 
except for the possibility of a wrong sign limit \cite{Ferreira:2014naa, Ferreira:2014dya},  
where the couplings to down-type quarks can have a relative sign
to the gauge bosons ones, thus oppositely to those of the SM. Our benchmarks will focus
on the SM-like limit where indeed $\cos(\beta-\alpha) \approx 0$.

We end this section by noticing that we have used the public program 2HDMC \cite{Eriksson:2009ws} to evaluate the 2HDM spectrum as well as the decay rates and BRs of all Higgs particles. {{{We have used 2HDMC to also enforce the aforementioned theoretical constraints onto both BSM scenarios considered here}}.}.

\section{Theoretical and experimental constraints}
In order to perform a systematic scan over the parameter space of the two MSSM configurations and the 
four 2HDM Types,  we take into account the following theoretical\footnote{Notice that, for  the MSSM scenarios considered here, the (dynamically generated) scalar potential is  stable in vacuum and does not induce perturbative unitarity violations.} and experimental constraints. 

\subsection{Theoretical constraints}

We list these here as itemized entries.
\begin{itemize}
	\item \textbf{\underline{Vacuum stability}}
	To ensure that the scalar potential is bounded from below,  it is enough to assume that 
	the quartic couplings should satisfy the following  relations~\cite{Deshpande:1977rw}:
	\begin{equation}
	\lambda_{1,2}>0,\qquad \lambda_3>-(\lambda_1 \lambda_2)^{1/2}\qquad \mathrm{and} \qquad \lambda_3+\lambda_4- \vert \lambda_5 \vert  >-(\lambda_1 \lambda_2)^{1/2}.
	\end{equation}		
	We also impose that the potential has a minimum that is compatible with EWSB.
	If this minimum is CP-conserving, any other possible charged or CP-violating stationary points
	will be a saddle point above the minimum~\cite{Ferreira:2004yd}. However, there is still the possibility
	of having two coexisting CP-conserving minima. In order to force the minimum compatible with
	EWSB, one need to impose the following simple condition~\cite{Barroso:2013awa}:
	\begin{equation}
	m_{12}^2 \left(m_{11}^2-m_{22}^2 \sqrt{\lambda_1/\lambda_2} \right) \left( \tan \beta - \sqrt[4]{\lambda_1/\lambda_2}\right) >0.
	\end{equation}

	Writing the minimum conditions as
	\begin{align}
	m_{11}^2+\dfrac{\lambda_1 v_1^2}{2}+\dfrac{\lambda_3 v_2^2}{2} &= \frac{v_2}{v_1} \left[ m_{12}^2 - (\lambda_4+\lambda_5)\dfrac{v_1 v_2}{2}\right],\nonumber\\
	m_{22}^2+\dfrac{\lambda_2 v_2^2}{2}+\dfrac{\lambda_3 v_1^2}{2} &= \frac{v_1}{v_2} \left[ m_{12}^2 - (\lambda_4+\lambda_5)\dfrac{v_1 v_2}{2}\right],
	\end{align} 
	allows us to express $m_{11}^2$ and $m_{22}^2$ in terms of the soft
	$Z_2$ breaking term $m_{12}^2$ and the quartic couplings $\lambda_{1-5}$.
	\item \textbf{\underline{Perturbative unitarity}}
	Another important theoretical constraint on the (pseudo)scalar sector of  the 2HDM  is
	the perturbative unitarity requirement. We require that the $S$-wave component 
	of the various (pseudo)scalar scattering {amplitudes} of Goldstone and Higgs {states remains unitary}.
	Such a condition implies a set of constraints that have to 
	be fulfilled and are given by~\cite{Akeroyd:2000wc}
	\begin{equation}
	|a_\pm|, |b_\pm|, |c_\pm|, |f_\pm|, |e_{1,2}|, |f_1|, |p_1| < 8 \pi,
	\end{equation}
	where
	\begin{align}
	\begin{split}
	a_\pm &= \dfrac{3}{2}(\lambda_1+\lambda_2)\pm \sqrt{\dfrac{9}{4}(\lambda_1-\lambda_2)^2+(2\lambda_3+\lambda_4)^2},\\
	b_\pm &= \dfrac{1}{2}(\lambda_1+\lambda_2)\pm \dfrac{1}{2} \sqrt{(\lambda_1-\lambda_2)^2+4\lambda_4^2},\\
	c_\pm &= \dfrac{1}{2}(\lambda_1+\lambda_2)\pm \dfrac{1}{2} \sqrt{(\lambda_1-\lambda_2)^2+4\lambda_5^2},\\
	e_1 &= \lambda_3 + 2 \lambda_4 -3\lambda_5,\hspace*{3cm}
	e_2 = \lambda_3-\lambda_5,\\
	f_+ &= \lambda_3+2 \lambda_4+3\lambda_5, \hspace*{2.9cm} f_- =\lambda_3+\lambda_5,\\
	f_1 &= \lambda_3+\lambda_4, \hspace*{4.3cm}p_1 = \lambda_3-\lambda_4.
	\end{split}
	\end{align}
	\item \textbf{\underline{EW Precision Observables (EWPOs)}}
	The additional neutral and charged (pseudo)scalars, beyond the SM-like Higgs state,  contribute to the gauge bosons vacuum
	polarization through their coupling to gauge bosons. 
	In particular, the universal parameters
	$S$, $T$ and $U$ provide constraints on the mass splitting 
	between the heavy states $m_H$, $m_{H^\pm}$ and $m_A$
	in the scenario in which $h$ is identified with the SM-like Higgs state. 
	The general expressions for the parameters $S$, $T$ and $U$ in 2HDMs  
	can be found in~\cite{Kanemura:2015mxa}. 
	To derive constraints on the scalar spectrum we consider the following 
	values for $S, T$ and $U$:
	\begin{equation}
	\Delta S = 0.05\pm 0.11,\quad \quad  \Delta T = 0.09\pm 0.13, \quad \quad 
	\Delta U = 0.01\pm 0.11, 
	\end{equation}
	while using the corresponding covariance matrix given in \cite{Baak:2014ora}.
	The $\chi^2$ function is then expressed as
	\begin{equation}
	\chi^2 = \sum_{i,j}(X_i - X_i^{\rm SM})(\sigma^2)_{ij}^{-1}(X_j - X_j^{\rm SM}),
	\end{equation}
	with correlation factor  +0.91. 
\end{itemize}

The aforementioned 2HDMC program  allows us to check most of  the above theoretical constraints, such as perturbative unitarity, boundedness from below of the scalar potential as well as EWPOs ($S, T$ and $U$),  
which are all turned on during the calculation, and can be adapted to the MSSM as well.

\subsection{Experimental constraints}
The parameter space of our benchmark scenarios is already partially
constrained by the limits obtained from various searches for
additional Higgs bosons at the LHC and elsewhere as well as  the requirement that one of
the neutral scalar states should match the properties of the observed SM-like Higgs 
boson. 
We evaluate the former constraints with the  
code {HiggsBounds}~\cite{Bechtle:2008jh,
	Bechtle:2011sb,Bechtle:2013wla,Bechtle:2015pma} 
and the latter with the code 
{HiggsSignals}~\cite{Bechtle:2013xfa}.
We stress, however, that our study of the existing constraints
cannot truly replace a dedicated analysis of the proposed benchmark
scenarios by ATLAS and CMS, which alone would be able to combine the
results of different searches taking into account all correlations.
In this section we briefly summarize the relevant features of  
{HiggsBounds} and {HiggsSignals} used in our study.

\subsubsection{Collider constraints }
The code {HiggsBounds} tests each parameter point for
$95\%$ Confidence Level (CL) exclusion from Higgs searches at the LHC as well as LEP
and Tevatron. 
First, the code determines the most sensitive experimental
search available, as judged by the expected limit,  for each
additional Higgs boson in the model. Then, only the selected channels
are applied to the model, i.e., the predicted signal rate for the most
sensitive search of each additional Higgs boson is compared to the
observed upper limit. In the case the prediction exceeds the limit, the
parameter point is regarded as excluded. For more details on the
procedure, the reader can see Ref. \cite{Bechtle:2015pma}.

Among the searches that are
relevant in constraining our scenarios for charged Higgs studies, 
the version we have used, {5.2.0beta}, of {HiggsBounds} includes the following.
\begin{itemize}
	\item ATLAS~\cite{Aaboud:2017sjh} and CMS~\cite{Sirunyan:2018zut} searches
	for heavy Higgs bosons decaying to $\tau^+\tau^-$ pairs using about
	$36\,\mathrm{fb}^{-1}$ of Run-2 data as well as the CMS results from
	Run-1~\cite{CMS:2015mca}.
	
	\item Searches at Run-1 and Run-2  by
	ATLAS~\cite{Aad:2015kna, Aaboud:2017rel} and
	CMS~\cite{Khachatryan:2015cwa, Sirunyan:2018qlb} for a heavy scalar
	decaying to a $Z$ boson pair, $H\to ZZ$. 
	
	\item Searches at Run-1 and Run-2 by
	ATLAS~\cite{Aad:2015xja} and 
	CMS~\cite{Sirunyan:2017djm, Sirunyan:2017guj} for a heavy scalar
	decaying to a pair of $125$ GeV SM-like Higgs scalars, $H\to hh$. 
	
	\item Searches at Run-1  by
	ATLAS~\cite{Aad:2015bua} and CMS~\cite{Khachatryan:2017mnf} for the
	$125$ GeV scalar decaying to a pair of lighter pseudoscalars, $h\to AA$. 
	
	\item Searches at Run-1 by ATLAS~\cite{Aad:2015wra} and
	CMS~\cite{Khachatryan:2015lba} for a heavy pseudoscalar decaying to a
	$Z$ boson and the 125 GeV scalar, $A\to Zh$.
\end{itemize}

By comparing these results with the predictions of SusHi, 
FeynHiggs and 2HDMC for the production cross sections and decay BRs of the
additional neutral Higgs bosons, {HiggsBounds} reconstructs
the 95\% CL~exclusion contours for our benchmark scenarios.  
In the MSSM and 2HDM Type II,  these constraints are typically stronger for large values of $\tan\beta$,
due to an enhancement of the production cross section of the heavier
Higgs bosons in bottom-quark annihilation (in that case the most
relevant searches are those for the decay to a $\tau^+\tau^-$ pair).
However, this is not generally true in the other 2HDM Types.
{HiggsBounds} also contains the available constraints from 
searches for a charged Higgs boson by ATLAS and CMS. Most relevant in
our scenarios are the constraints on the production of a light charged
Higgs via top quark decay, $t\to H^+ b$, with subsequent decay
$H^+ \to \tau^+ \nu$~\cite{Aad:2014kga,Khachatryan:2015qxa, CMS:2016szv, Aaboud:2018gjj}, as well as top-quark associated
$H^\pm$ production, with subsequent decays to the
$\tau\nu$~\cite{Aad:2014kga,Khachatryan:2015qxa, CMS:2016szv, Aaboud:2018gjj} and/or $tb$  \cite{Khachatryan:2015qxa, ATLAS:2016qiq, Aad:2015typ} channels.

In order to estimate the theoretical uncertainty in our
determination of the excluded regions, we rely on the uncertainty
estimates for the gluon-fusion and bottom-quark annihilation cross sections. 
The most
conservative (i.e., weakest) determination of the exclusion region is
obtained by taking simultaneously the lowest values in the uncertainty
range for both production processes of each of the heavier Higgs
bosons, while the least conservative (i.e., strongest) determination
is obtained by taking simultaneously the highest values in the
uncertainty range.

With the use of the code HiggsSignals, we  test the compatibility of our
scenarios with the observed SM-like Higgs signals, by comparing the predictions
of SusHi, FeynHiggs and 2HDMC for the signal strengths of Higgs production and
decay in a variety of channels against ATLAS and CMS  measurements.
The version we have used, {2.2.0beta}, of {HiggsSignals} includes all
the combined ATLAS and CMS results from Run-1 of the
LHC~\cite{Khachatryan:2016vau} as well as all the available
ATLAS~\cite{ATLAS:2016gld, ATLAS:2018gcr, Aaboud:2017xsd,
	Aaboud:2017vzb, Aaboud:2017jvq, Aaboud:2017rss, Aaboud:2018xdt} and
CMS limits from Run-2~\cite{CMS:2017rli, Sirunyan:2017exp, Sirunyan:2017khh,Sirunyan:2017dgc, 
	Sirunyan:2017elk, Sirunyan:2018shy, Sirunyan:2018ygk, Sirunyan:2018mvw, Sirunyan:2018egh}.

\subsubsection{DM constraints}

{{These have naturally been enforced only in the MSSM case, by
    using the program micrOMEGAs version 5.0.9 \cite{Belanger:2013oya,mO}. 
Such a code  calculates the properties of DM in terms of its 
relic density as well as its direct and  indirect detection rates. 
For the two MSSM scenarios considered here, the DM candidate, i.e., 
the Lightest Supersymmetric Particle (LSP),  is the lightest neutralino. 
We require that the outcome of the calculation of the relic density 
should be in agreement with the latest Planck measurement 
\cite{Aghanim:2018eyx}.}}

\section{Numerical results}
\label{sec:results-MSSM}
In this  section, we present our findings for the MSSM and 2HDM in turn.

\subsection{MSSM results} \label{sec:first scenario}
In the hMSSM scenario, all superparticles are chosen to be rather heavy so that production and
decays of the MSSM Higgs bosons are only mildly affected by their
presence due to the decoupling properties of SUSY. 
In particular, the loop-induced SUSY contributions to the
couplings of the light CP-even scalars are small and the heavy
Higgs bosons with the masses even up to $2$~TeV decay only to SM particles.
Therefore, the phenomenology of this scenario at the LHC resembles that of a 2HDM Type-II 
with MSSM-inspired Higgs couplings and mass relations. 
%

The masses of the third generation squarks and that of the gluino are
safely above the current bounds from direct searches at the LHC, as intimated. Specifically, we refer to
\cite{Sirunyan:2017kqq, Sirunyan:2018vjp, Aaboud:2017nfd,
	Aaboud:2017ayj, Aaboud:2017aeu, Sirunyan:2017xse, Sirunyan:2017wif,
	Sirunyan:2017leh} for the scalar top quarks,
\cite{Sirunyan:2017kqq, Sirunyan:2018vjp, Aaboud:2017phn, Aaboud:2017dmy, Aaboud:2017wqg} for the 
scalar bottom quarks and \cite{Sirunyan:2018vjp, Aaboud:2017dmy,Aaboud:2017bac, Aaboud:2017vwy, Sirunyan:2017cwe} for the gluino. 
The value chosen for $X_t$ is close to the
one for which the maximal value of $m_h$ is obtained.  
The $m^{\rm mod+}_h$  scenario is very similar to the hMSSM one except the fact that we take  
$X_t=2M_{\rm SUSY}=1$ TeV.

\begin{figure}[h!]
	\centering
	\includegraphics[width=7cm, height = 7.5cm]{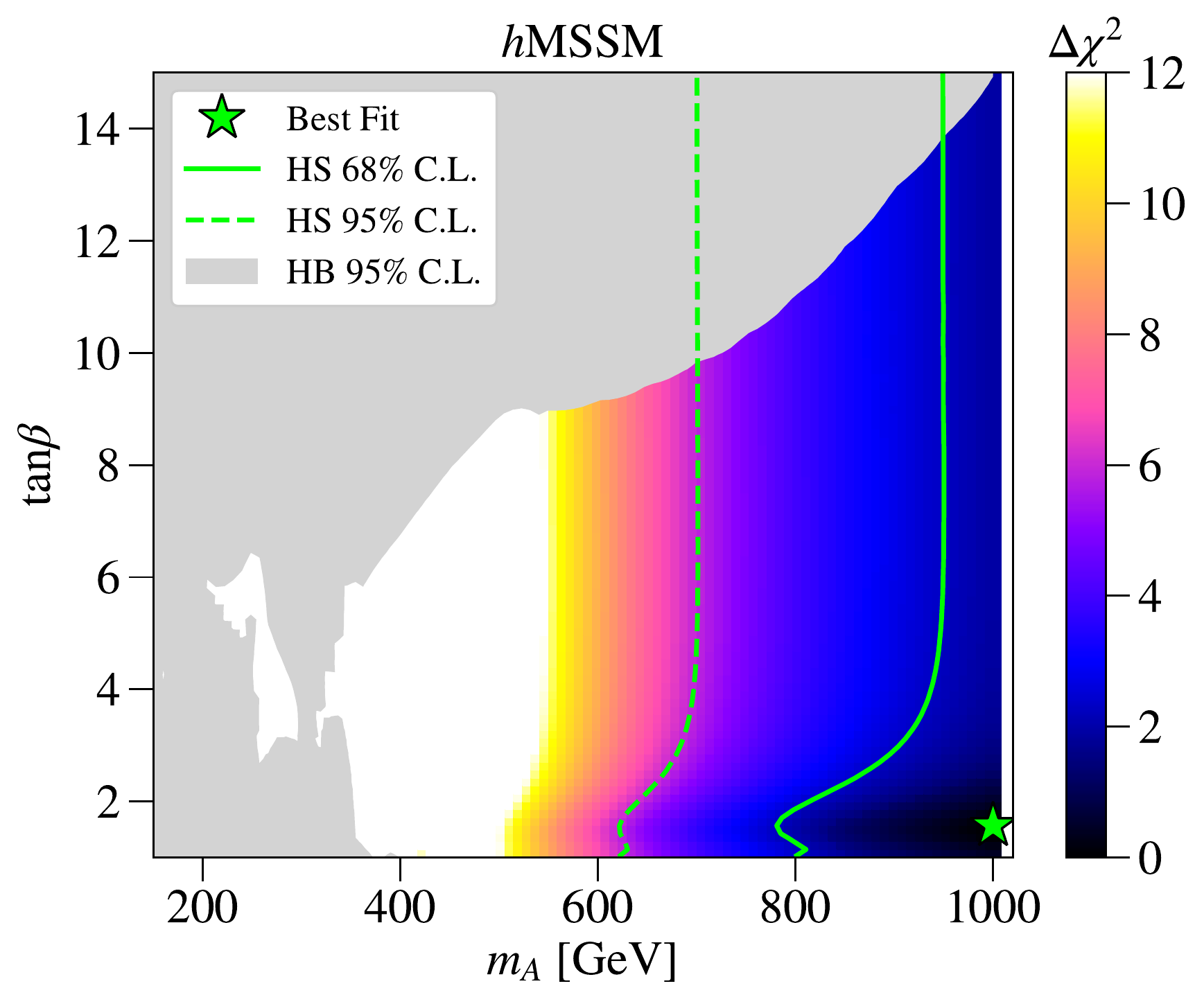}
	\includegraphics[width=7cm, height = 7.5cm]{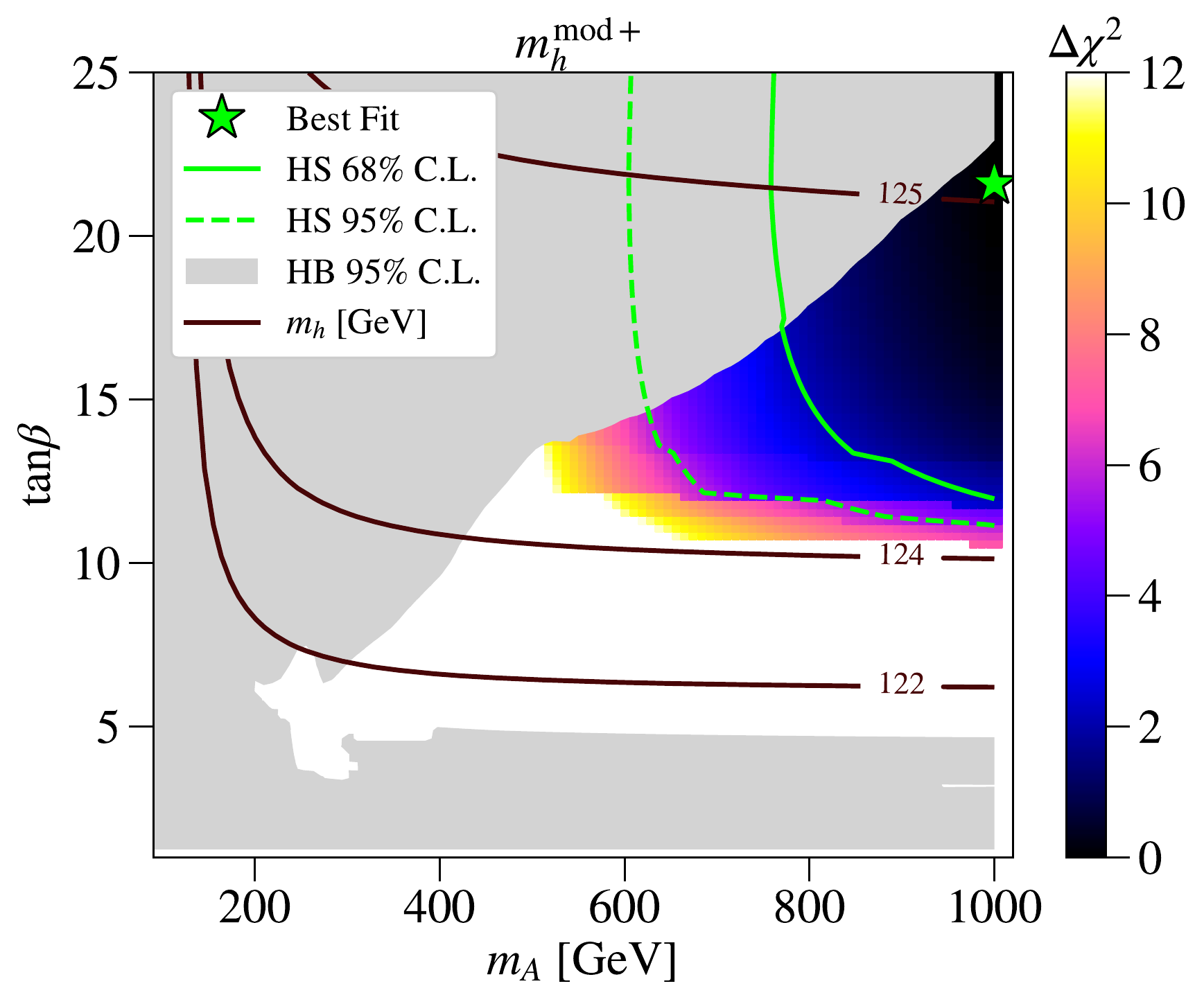}
	\vspace{0.7cm}
	\caption{The allowed regions on the ($m_A, \tan\beta$) plane in hMSSM (left)
		and $m^{\rm mod+}_h$ (right). The cyan lines in the right plot are level curves for the SM-like Higgs mass. 
		{{By definition, in the hMSSM,  $m_h$ is fixed at 125 GeV. The best
fit points are marked by green stars.}}}
	\label{mssm:fig1}
\end{figure}
\begin{figure}[h!]
	\centering
	\includegraphics[width=7cm, height = 7cm]{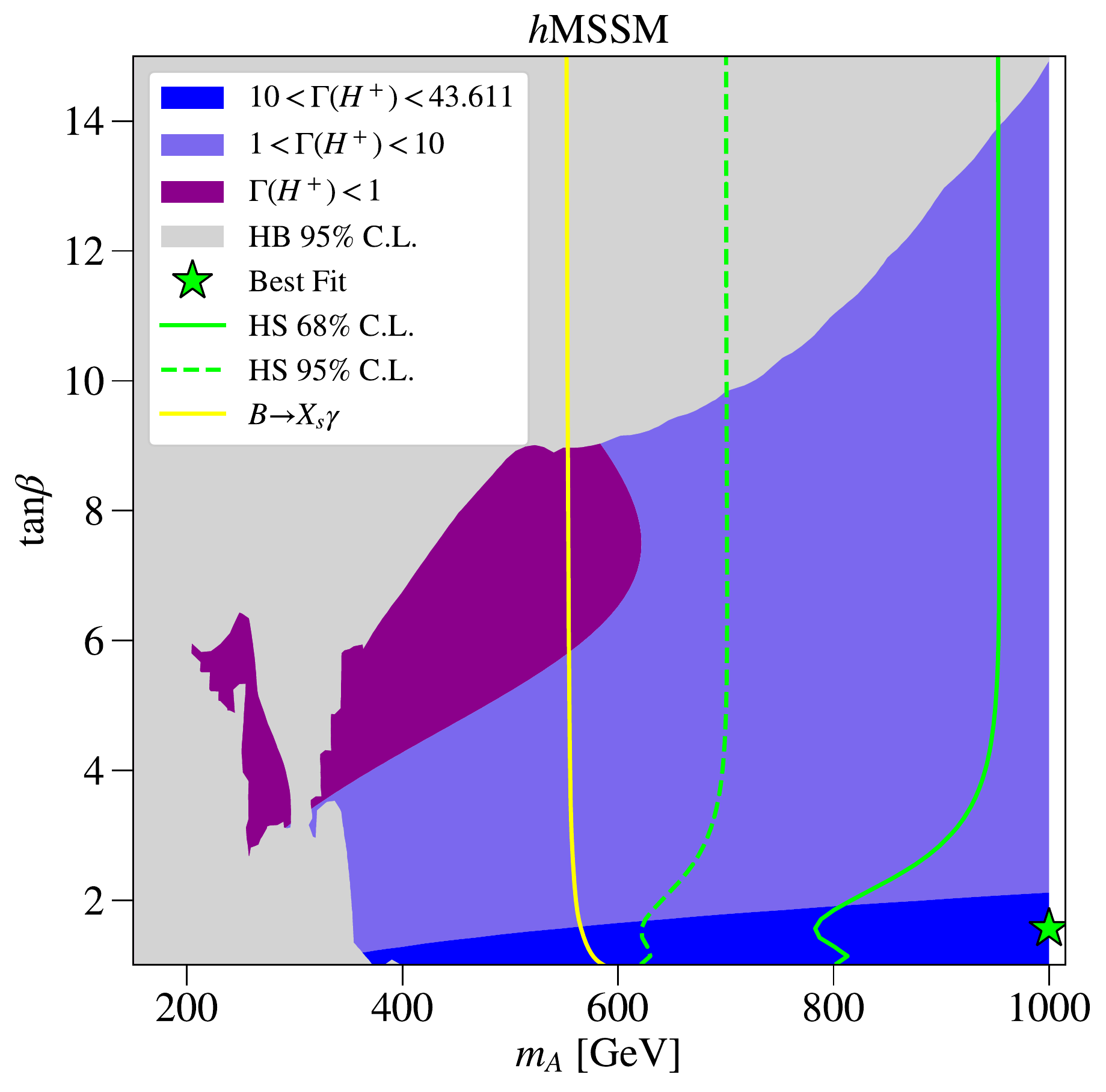}
	\includegraphics[width=7cm, height = 7cm]{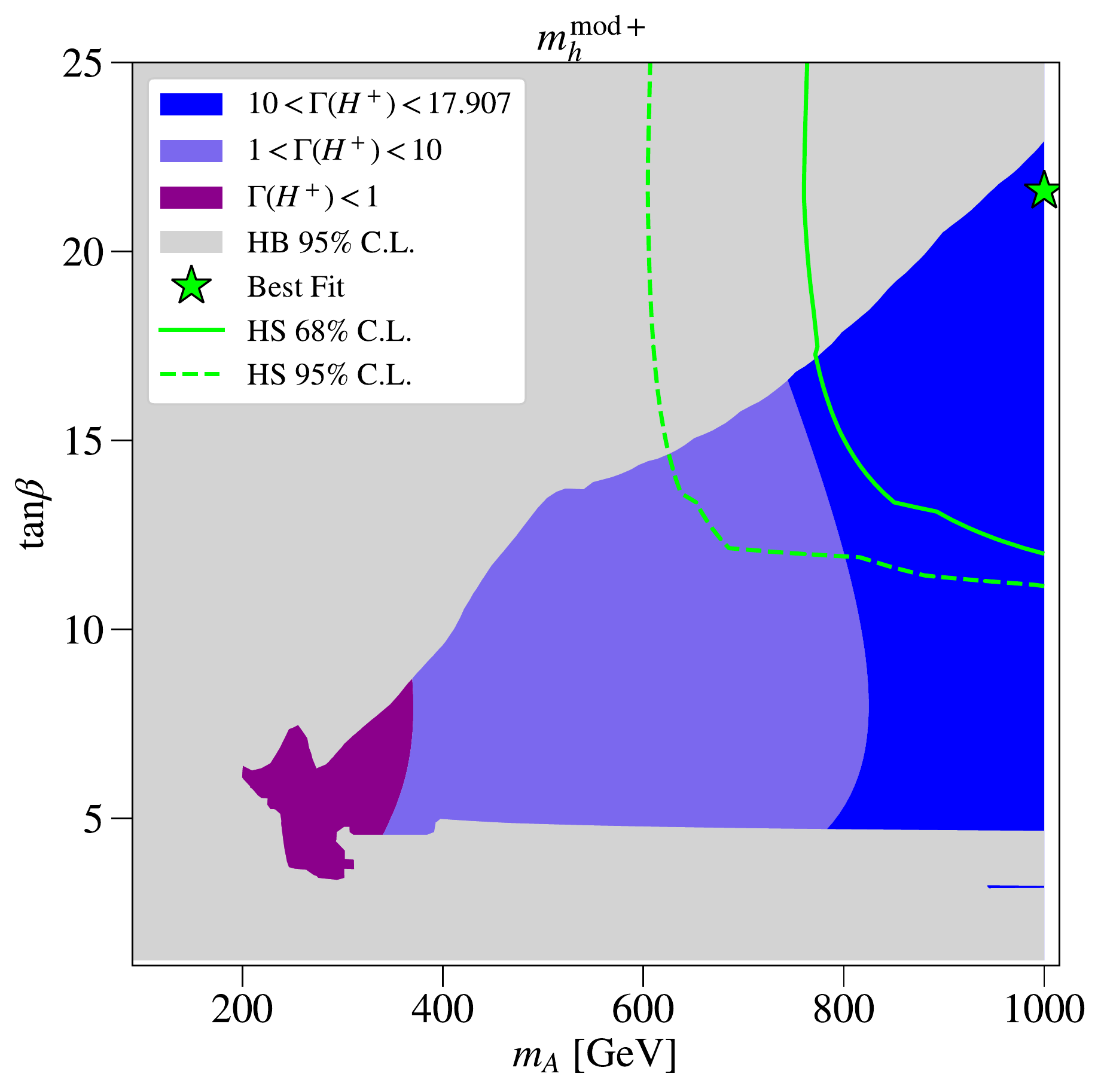}
	\vspace{0.7cm}
	\caption{Total charged Higgs boson width (in GeV) mapped on the ($m_A, \tan\beta$) plane in hMSSM (left)
		and $m^{\rm mod+}_h$ (right). The units of $\Gamma_{H^\pm}$ in the legends are intended in GeV.
           The best fit points are marked by green stars.}
	\label{mssm:fig2}
\end{figure}

In Fig. \ref{mssm:fig1} the allowed regions on the ($m_A, \tan\beta$) plane are depicted for various $\Delta\chi^2$, wherein the  
left and right panel are,  respectively, for the hMSSM and $m^{\rm mod+}_h$ scenarios. 
For the hMSSM and $\Delta\chi^2\leq 12$, one can see that $m_A$ should be heavier than about 400 GeV.
In the case of  $m_A\approx 400-600$ GeV, $\tan\beta$ should be in the range $[1,9]$ while for $m_A$ around 
1 TeV $\tan\beta$  is in the range $[1,15]$. The dashed(solid) line represents the 95\%(68\%) CL obtained by the HiggsSignals fit and the best fit point is located at $m_A\approx 1$ TeV and $\tan\beta \approx 2$.
For the $m^{\rm mod+}_h$ scenario, the situation is quite different. 
In order to {accommodate} $m_h\approx 125$ GeV, one needs $\tan\beta>10$. Similarly to the left panel, also in the right one 
the dashed(solid) line represents the 95\%(68\%) CL obtained by the HiggsSignals fit and the best fit point is located at $m_A\approx 1$ TeV and $\tan\beta \approx 20$. For this scenario and for $\Delta\chi^2< 12$,
all $\tan\beta\leq 10 $ are excluded. {Note that after imposing DM constraints onto the best fit analysis, we observe that the best fit points for both the  hMSSM and $m^{\rm mod+}_h$ scenarios move to somewhat lighter values of the charged Higgs boson mass. The BPs given in Tab.~\ref{tab:benchmarks_points} account for this effect.}

\begin{figure}[h!]
	\includegraphics[width=8cm, height = 8cm]{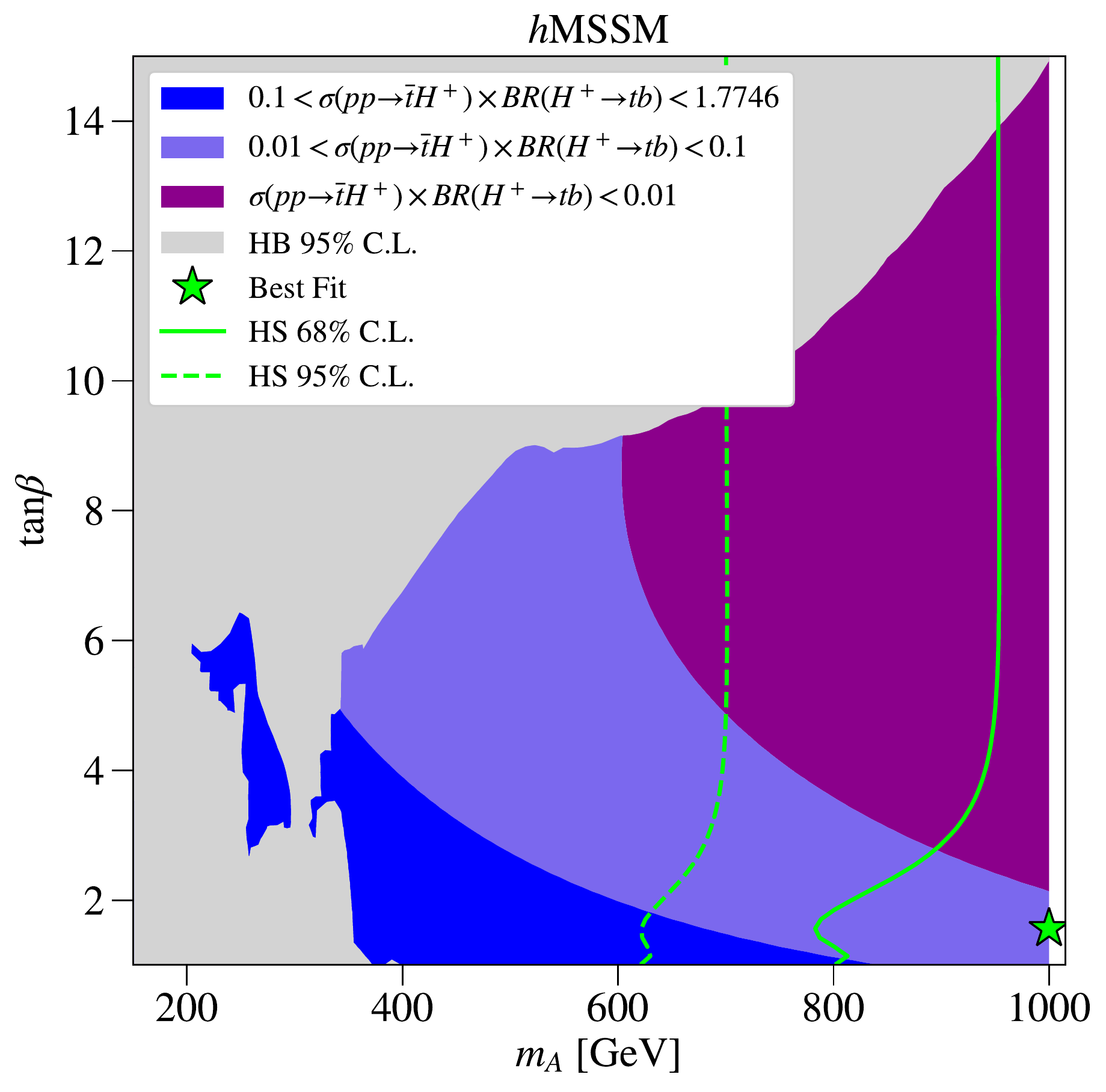}\includegraphics[width=8cm, height = 8cm]{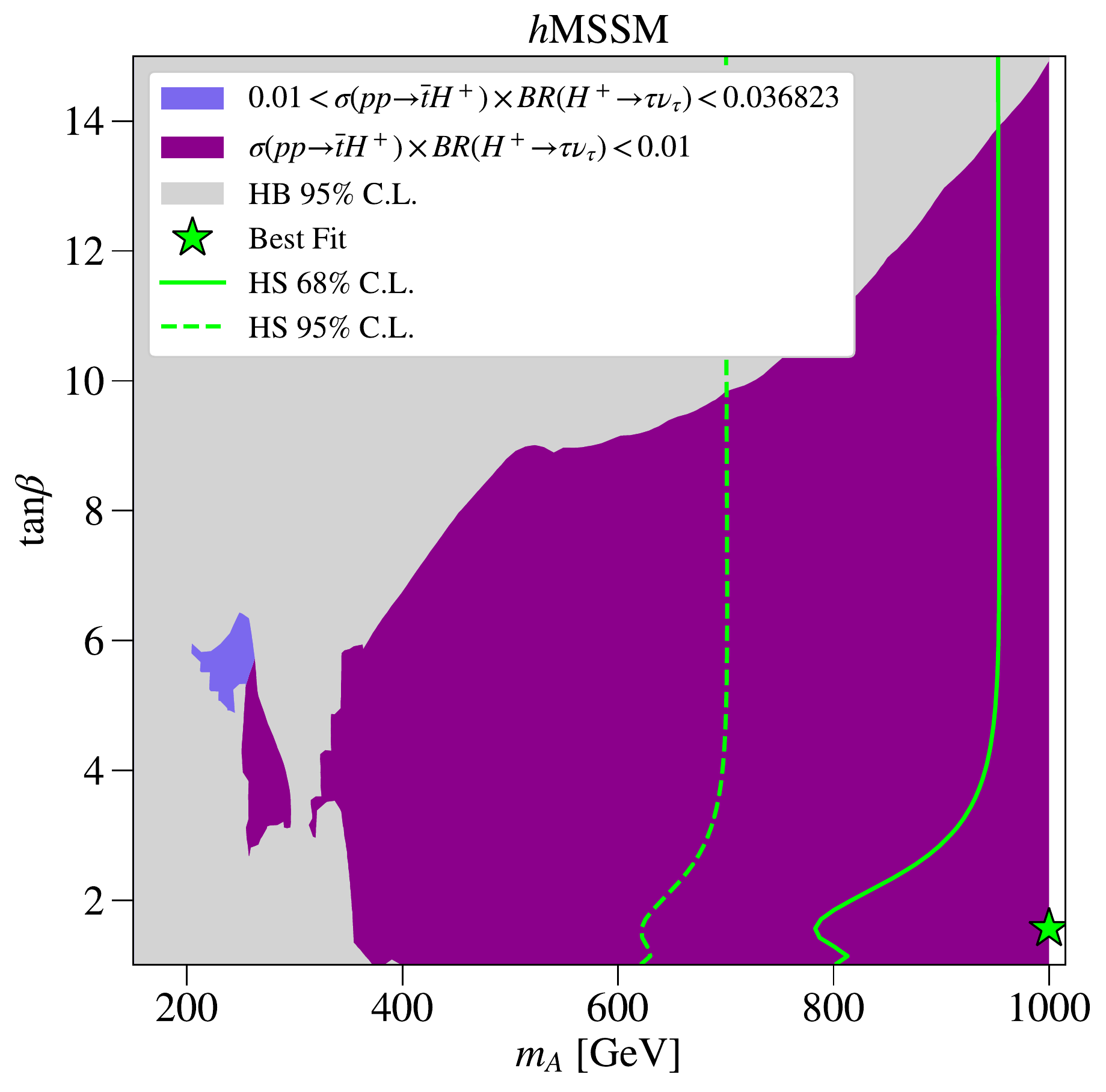}
	\includegraphics[width=8cm, height = 8cm]{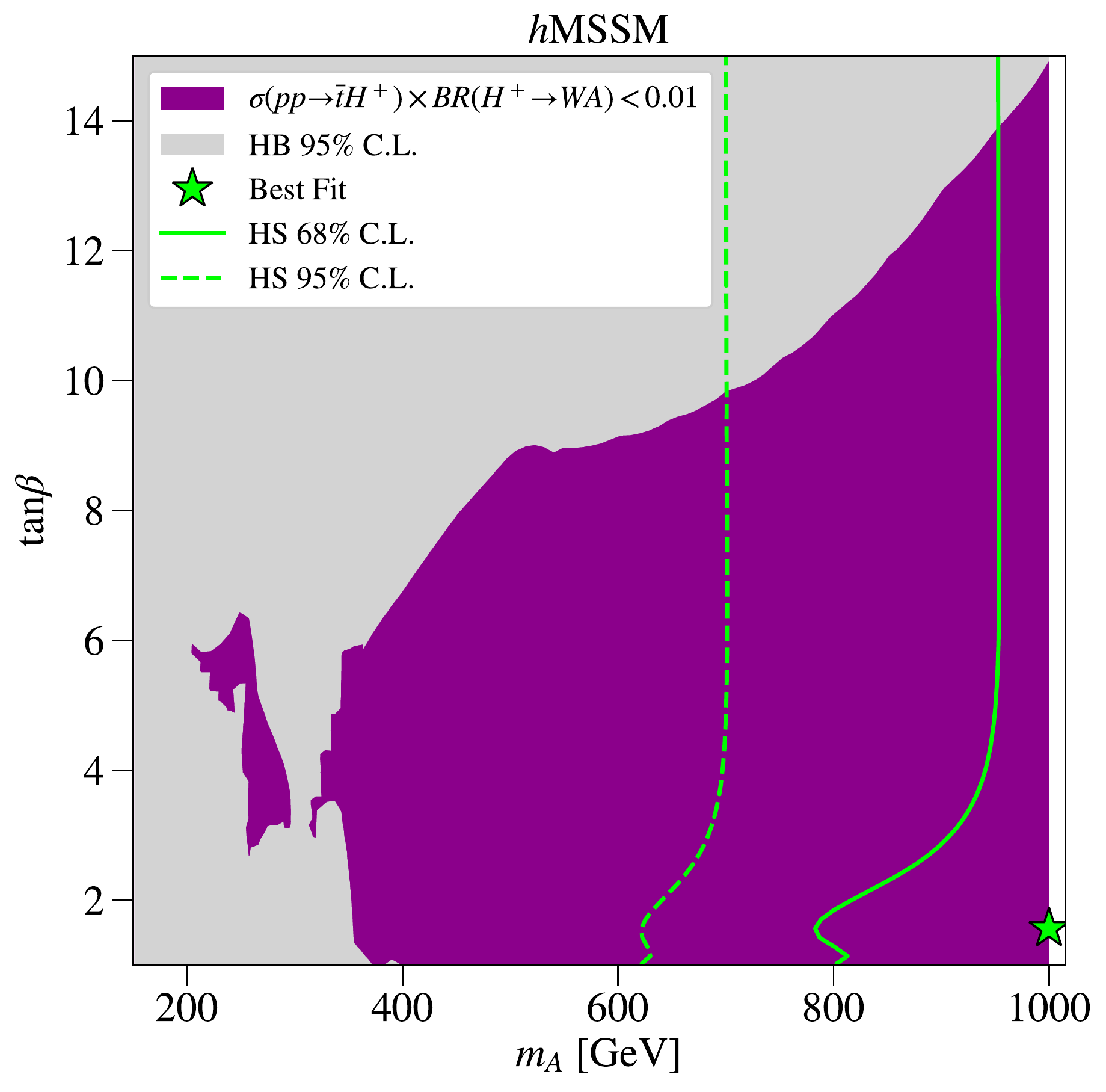}\includegraphics[width=8cm, height = 8cm]{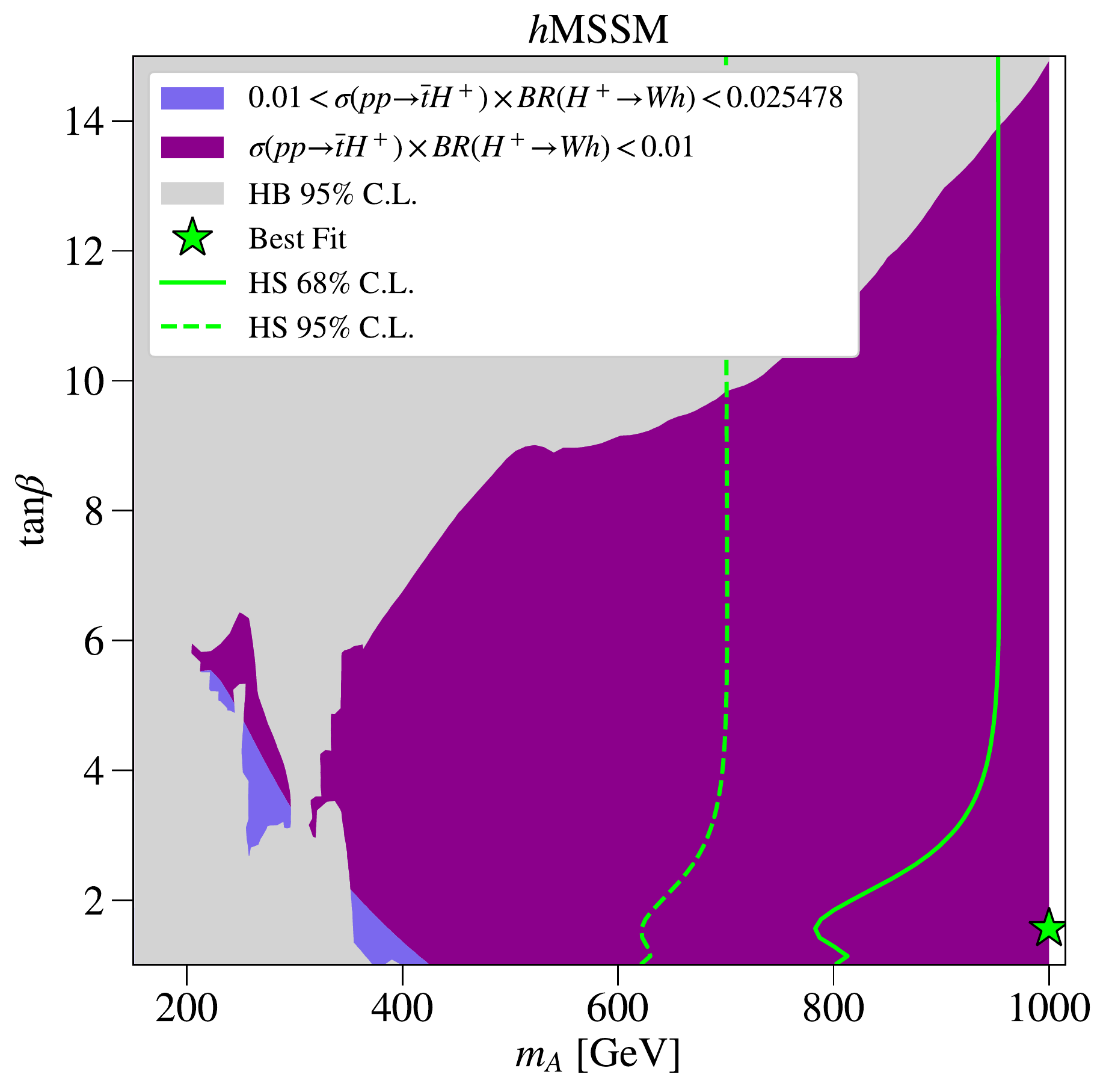}
	\vspace*{0.5truecm}
	\caption{The $\sigma(pp\to tH^-+{\rm c.c.})\times {\rm BR}(H^\pm\to XY)$ rate (in pb) at $\sqrt{s}$=14 TeV in the hMSSM scenario, for $XY\equiv tb$ (top left), $XY\equiv \tau\nu$ (top right), $XY\equiv AW^\pm$ (bottom left) and $XY\equiv hW^\pm$ (bottom right). Notice that c.c. channels are included. The best
fit points are marked by green stars.
	}
	\label{figure:hmssm}
\end{figure}

\begin{figure}[h!]
	
	\includegraphics[width=8cm, height = 8cm]{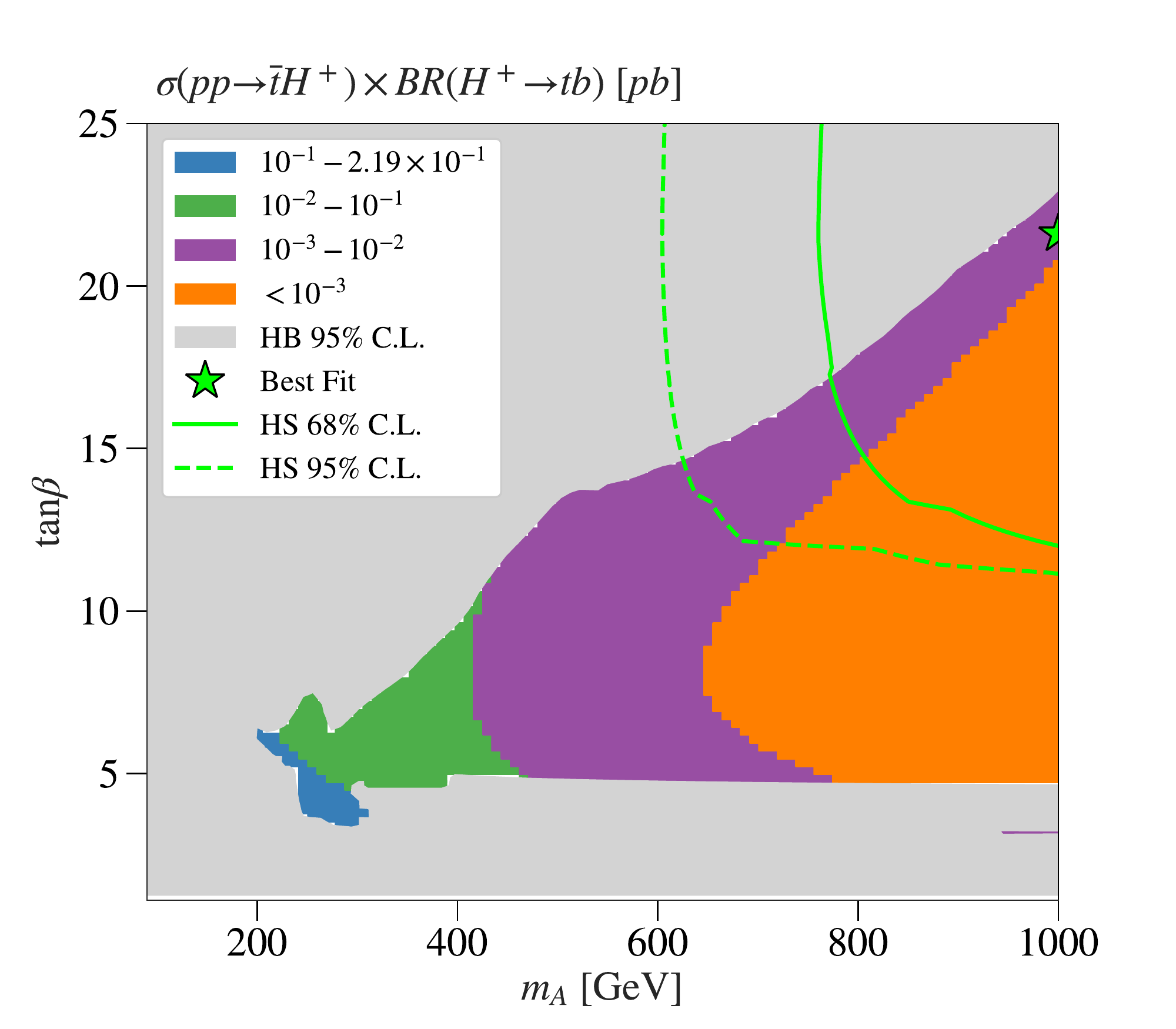} 
	\includegraphics[width=8cm, height = 8cm]{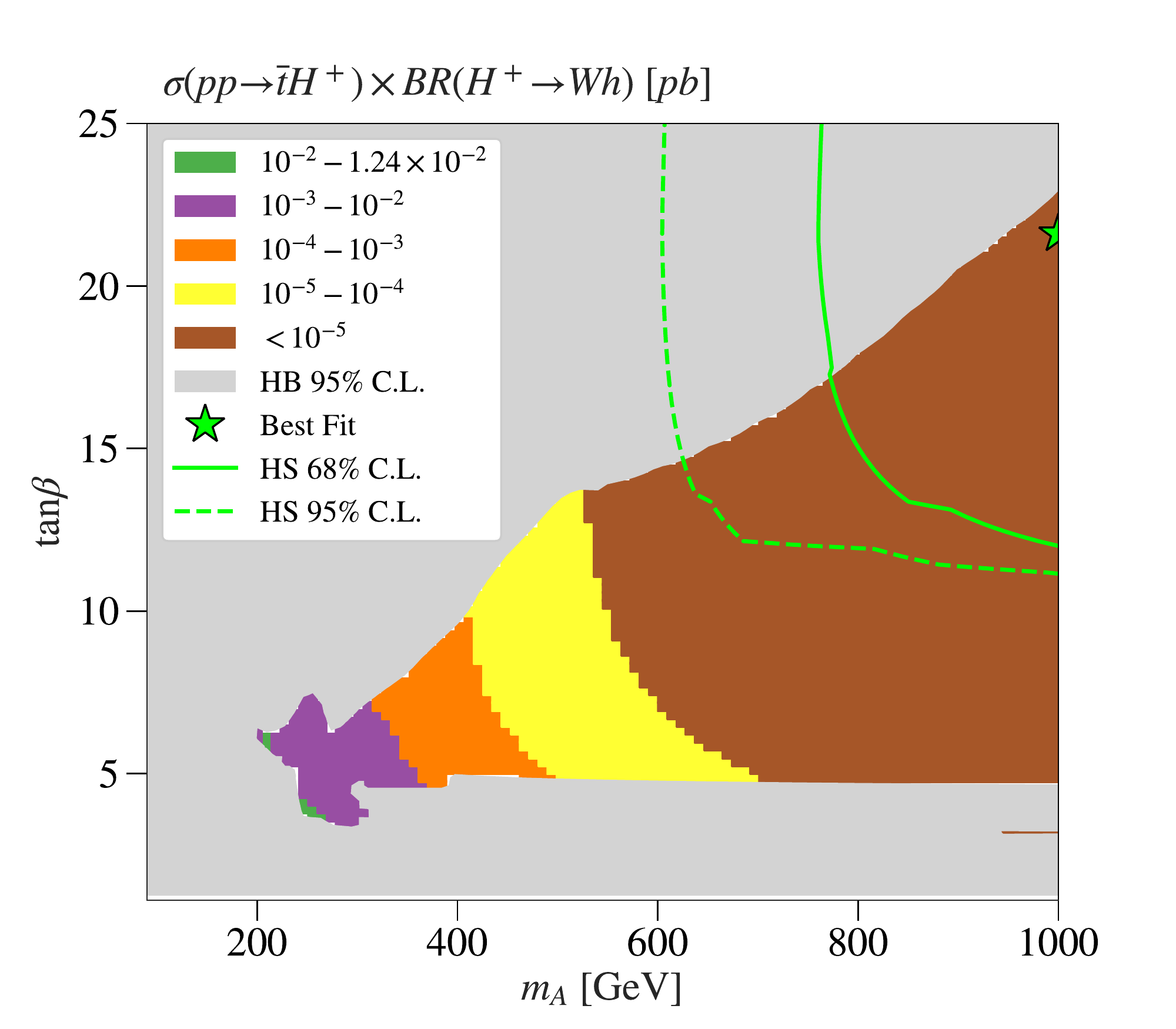} 
	
	\includegraphics[width=8cm, height = 8cm]{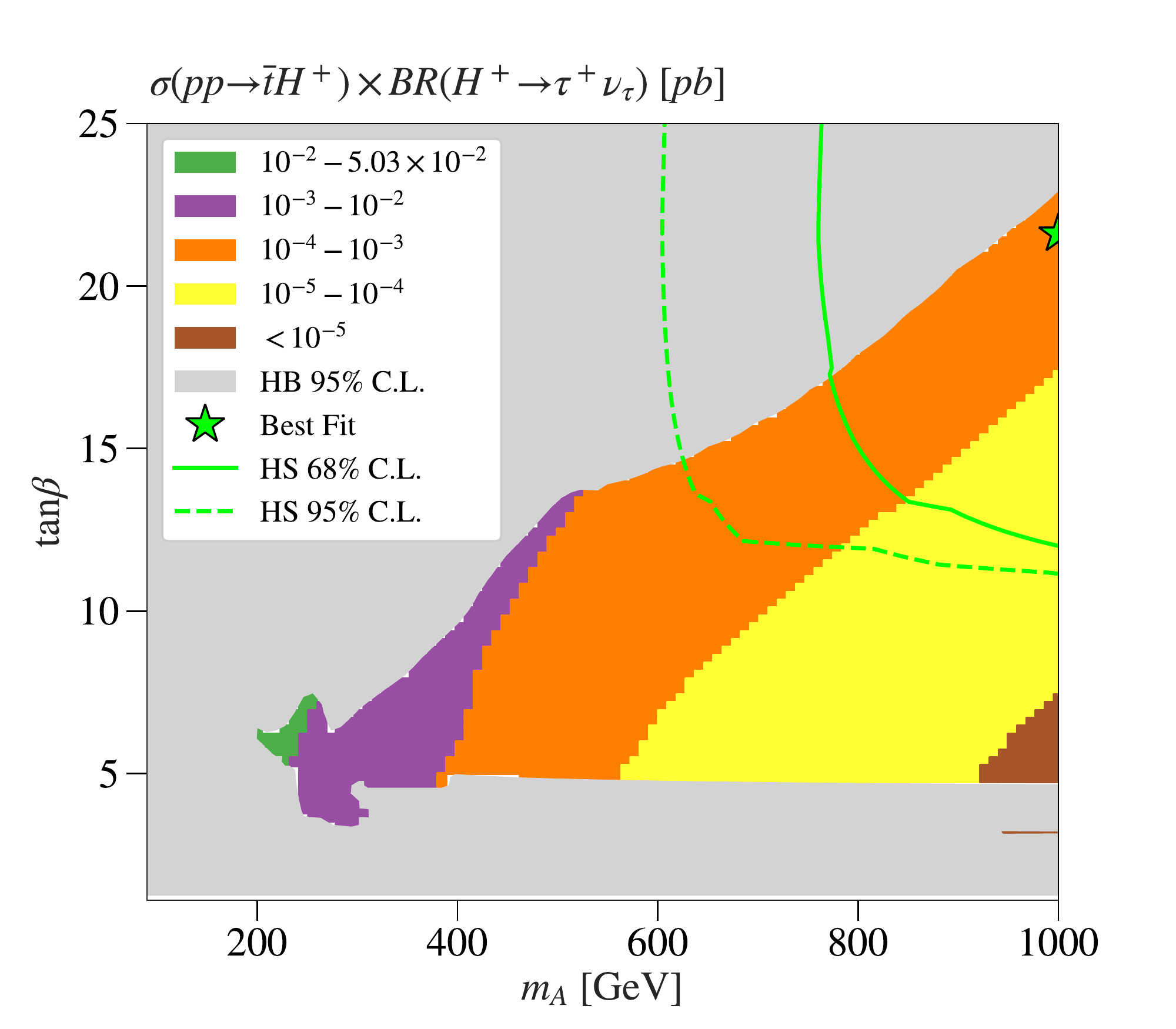}
	\includegraphics[width=8cm, height = 8cm]{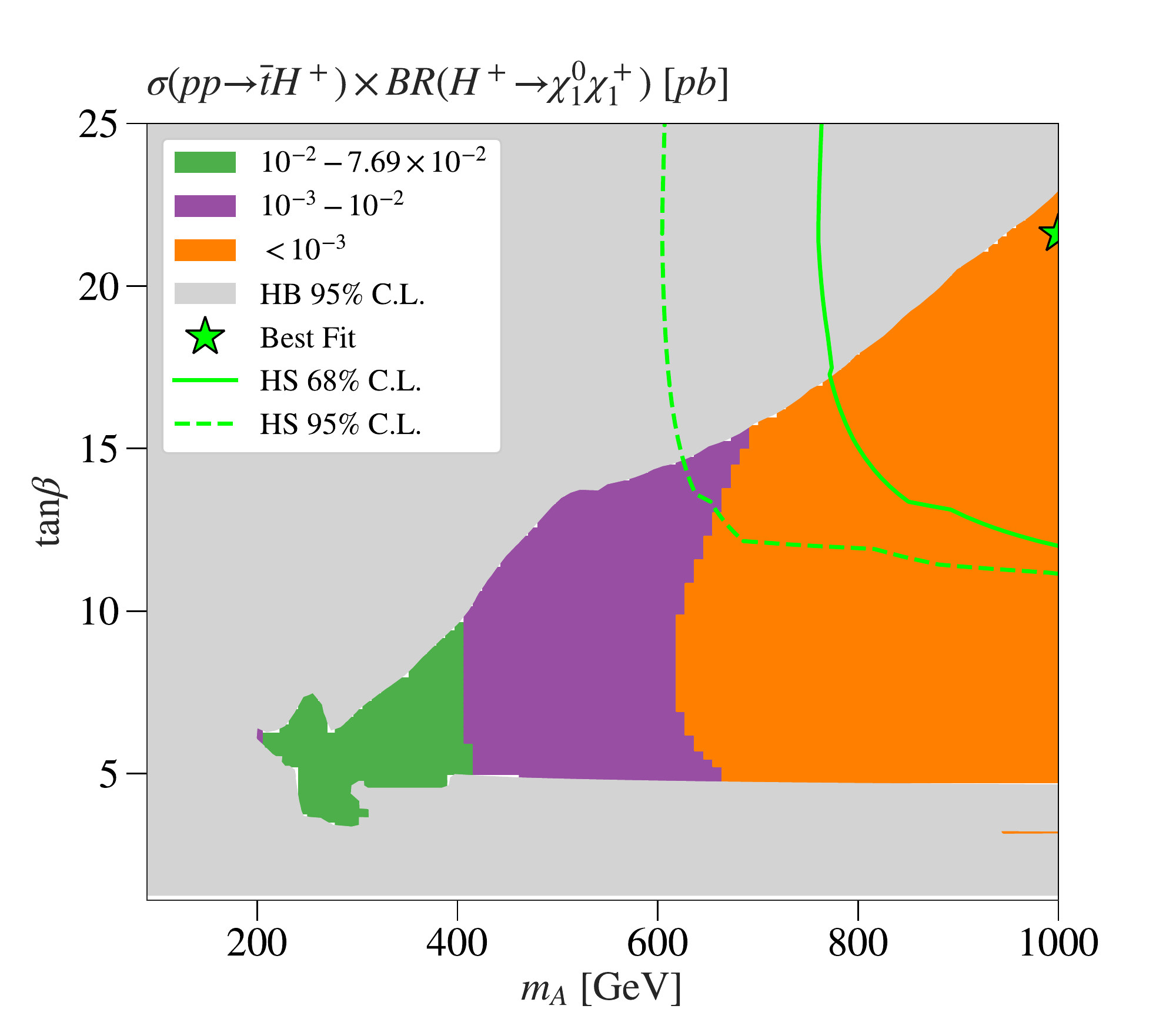}
	\vspace{0.5cm}
	\caption{The $\sigma(pp\to tH^-+{\rm c.c.})\times {\rm BR}(H^\pm\to XY)$ rate (in pb) at $\sqrt{s}$=14 TeV in the
		$m_{h}^{\rm mod+}$  scenario, for $XY\equiv tb$ (top left),
          $XY\equiv W^\pm h$ (top right), $XY\equiv \tau\nu$ (bottom left) and 
$XY\equiv \tilde{\chi}^0_1 \tilde{\chi}^+_1$ (bottom right). Notice that c.c. channels are included. 
	The best
fit points are marked by green stars.}
	\label{figure:mhmod}
\end{figure}

In Fig. \ref{mssm:fig2} we present  the total width of the charged Higgs boson,  again, over the $(m_A, \tan\beta)$ plane, 
for both hMSSM (left) and $m^{\rm mod+}_h$ (right). As one can see from the left panel, 
the total width for the hMSSM case is  largest for $\tan\beta\leq  3$, which is when  
$\Gamma_{H^\pm}\approx 7-10$ GeV, while for $\tan\beta\geq 5$
the width drops to 1--3 GeV. This effect can be attributed to the fact that the total width is fully dominated by 
$H^+ \to t\bar{b}$, whenever this channel is open, in which the top effect is more pronounced for low $\tan\beta$. In this case, $H^+ \to \tau \nu$ is subleading and also the decay modes $H^+\to \tilde{\chi}_i^+ \tilde{\chi}_j^0$ are suppressed.
In the case of $m^{\rm mod+}_h$, since small $\tan\beta$ is not allowed, the total charged Higgs boson width is generally smaller than
in the hMSSM case, as a consequence of the fact that $H^+\to t\bar{b}$ is therefore smaller in this scenario.
The maximal total width is here obtained for $m_A\approx 1$ TeV and a large $\tan\beta \approx 20$.
In the  $m^{\rm mod+}_h$ scenario, the decay $H^+ \to \tilde{\chi}_2^+ \tilde{\chi}_2^0$ could have a significant BR,   reaching 30\%. Hence, the $H^\pm$ is always rather narrow, whichever its mass. In fact, owing to the degeneracy between $m_A$ and $m_{H^\pm}$ in the MSSM, as dictated by $h$ data, a remarkable result is that in the minimal SUSY scenario a charged Higgs boson is essentially always heavier than the top quark.

In Fig. \ref{figure:hmssm} we show the production cross section for single charged Higgs boson production in association
with a top quark (as appropriate for the $m_{H^\pm}>m_t$ case) times the BR of $H^+$ into a specific final state for both the hMSSM and $m^{\rm mod+}_h$ scenarios using Prospino \cite{Dittmaier:2009np,Berger:2003sm,Beenakker:1996ed}. 
In fact, as we have seen previously, the total width of the charged Higgs state is rather small in both cases, in relation to the mass, so that  
one can use the Narrow Width Approximation (NWA) to estimate such a cross section (which we have done here).
In the top-left(top-right) panel of Fig.~\ref{figure:hmssm}, we show the size of the cross section of 
$\sigma(pp\to  tH^-+{\rm c.c.})\times {\rm BR}(H^\pm\to tb)$($\sigma(pp\to tH^-+{\rm c.c.})\times {\rm BR}(H^\pm\to \tau \nu)$), given in pb.

For the hMSSM scenario, one can see that in the $tb$ channel the largest cross section (more than 0.1 pb) 
is reached for small $\tan\beta < 3$.  There is also a wide region with 
$m_{H^\pm}\in [400,600]$ GeV and $\tan\beta < 10$ where the cross section is still rather important:
between $10^{-3}$ and 0.1 pb. 
As for the $\tau\nu$ channel, the cross section is maximized when $\tan\beta$ is in the range $[4,9]$ 
and the largest cross section is seen around $10^{-3}$ pb. However, amongst the bosonic channels,
$H^\pm\to W^\pm A$
is {hopeless} because BR$(H^\pm\to W^\pm A)$ is  very suppressed while $H^\pm\to W^\pm h$  
can have a rate that is close to $10^{-2}$ pb for small $\tan\beta\approx  1$. Note that, for completeness, we have also drawn 
the exclusion region due to  BR$(\overline{B}\to X_s \gamma)$, even though we can always assume some kind of flavor
violation that takes place in the MSSM and can bring the BR($\overline{B}\to X_s \gamma$) to a correct value.
In terms of $\sigma(pp\to  tH^-+{\rm c.c.})\times {\rm BR}(H^\pm\to XY)$ for the $m_{h}^{\rm mod+}$  scenario, the situation 
is worse. The best channels are $H^+\to t\bar{b}$ and   
$H^+ \to \tilde{\chi}_1^+ \tilde{\chi}_1^0$ with  
the maximum cross section in the allowed region being between   $10^{-3}$ and $10^{-2}$ pb
for charged Higgs boson masses in the range 400 to 600 GeV, as can be seen from Fig. \ref{figure:mhmod}.


\begin{table}[!t]
    \centering
    {\renewcommand{\arraystretch}{1}
    {\setlength{\tabcolsep}{0.5cm}
    \begin{tabular}{|c|c|c|}
        \hline
        Parameters & $h$MSSM & $m^{\rm mod+}_h$ \\
        \hline
        \hline
        \multicolumn{3}{|c|}{MSSM inputs} \\
        \hline
        $\tan\beta$ & $1.804$ & $5.9495$ \\
        $\mu \ \rm{(GeV)}$ & $200$ & $200$ \\
        $M_2 \ \rm{(GeV)}$ & $200$ & $200$ \\
        $m_{\tilde{g}} \ \rm{(GeV)}$ & $1500$ & $1500$ \\
        $A_t=A_b=A_\tau \ \rm{(GeV)}$ & $2110.9$ & $1533.6$ \\
        $M_{{Q}_{1,2}}=M_{{U}_{1,2}}=M_{{D}_{1,2}} \ \rm{(GeV)}$ & $1500$ & $1500$ \\
        $M_{{Q}_{3}}=M_{{U}_{3}}=M_{{D}_{3}} \ \rm{(GeV)}$ & $1000$ & $1000$ \\
        $M_{{L}_{1,2}}=M_{{E}_{1,2}} \ \rm{(GeV)}$ & $500$ & $500$ \\
        $M_{{L}_3}=M_{{E}_3} \ \rm{(GeV)}$ & $1000$ & $1000$ \\
        \hline
        \hline
        \multicolumn{3}{|c|}{Masses in GeV} \\
        \hline
        $M_{h^0}$ & $125$ & $118.45$ \\
        $M_{H^0}$ & $504.23$ & $222.35$ \\
        $M_{A^0}$ & $493.43$ & $218.69$ \\
        $M_{H^+}$ & $499.94$ & $232.91$ \\
        $M_{\tilde{b}_1}$ & $1109.7$ & $1000$ \\
        $M_{\tilde{b}_2}$ & $3041.3$ & $1002$ \\
        $M_{\tilde{t}_1}$ & $990.91$ & $876.49$ \\
        $M_{\tilde{t}_2}$ & $1230.4$ & $1134.8$ \\
        $M_{\tilde{\tau}_1}$ & $999$ & $1000.7$ \\
        $M_{\tilde{\tau}_2}$ & $1002.1$ & $1001.3$ \\
        $M_{\tilde{\chi}^0_1}$ & $74.736$ & $84.302$ \\
        $M_{\tilde{\chi}^0_2}$ & $139.94$ & $147.15$ \\
        $M_{\tilde{\chi}^0_4}$ & $282.57$ & $271.82$ \\
        $M_{\tilde{\chi}^+_1}$ & $123.97$ & $139.89$ \\
        $M_{\tilde{\chi}^+_2}$ & $278.48$ & $270.84$ \\
        \hline
        \hline
        \multicolumn{3}{|c|}{Total decay width  in GeV} \\
        \hline
        $\Gamma(H^+)$ & $6.7338$ & $0.16321$ \\
        \hline
        \hline
        \multicolumn{3}{|c|}{$BR(H^+ \to XY)$ in \%} \\
        \hline
        $BR(H^+ \to \tilde{\chi}^0_1 \tilde{\chi}^+_1)$ & $-$ & $27.93$ \\
        $BR(H^+ \to \tau^+ \nu_{\tau})$ & $0.05$ & $10.1$ \\
        $BR(H^+ \to W^+ h^0)$ & $1.04$ & $4.08$ \\
        $BR(H^+ \to b t)$ & $98.74$ & $57.65$ \\
        \hline
        \hline
        \multicolumn{3}{|c|}{Cross sections in pb} \\
        \hline
        $\sigma(pp \to tH^+ + {\rm c.c.})$ & $0.246$ & $0.204$ \\
        \hline
    \end{tabular}}}
    \caption{Benchmark points for the $h$MSSM and $m^{\rm mod+}_h$ scenarios.}
    \label{tab:benchmarks_points}
\end{table}		
We conclude this section by presenting in Tab. \ref{tab:benchmarks_points} two BPs, one each for the 
$m_h^{\rm mod+}$ and hMSSM scenarios, to aid future analyses of Run-2 (and possibly Run-3) data from the LHC. Notice that these BPs do not correspond to the best fit points in these two MSSM configurations, as the latter would yield too small cross sections\footnote{Probably accessible only at the High-Luminosity LHC \cite{Gianotti:2002xx}.}, owing to the very large charged Higgs mass involved (of order 1 TeV). Yet, the BPs presented correspond to rather large values of $m_{H^\pm}$, as dictated by the compatibility tests of the  $m_h^{\rm mod+}$ and hMSSM scenarios with current datasets, still giving production and decay rates (in one or more channels) potentially testable in the near future.
		
\subsection{2HDM results}

We now move on to discuss the 2HDM. In this scenario, we consider $h$ as being again the 125 GeV SM-like Higgs   
and vary the other six parameters  as indicated in Tab.~\ref{table2}.  
When performing the scan over the 2HDM parameter space, other than taking into account 
the usual LHC, Tevatron and LEP bounds  (as implemented in HiggsBounds and HiggsSignals)  as well as the theoretical ones (as  implemented in 2HDMC), we also have to consider flavor observables. In fact, unlike the MSSM, where potentially significant contributions to (especially) $B$-physics due to the additional Higgs states entering the 2HDM beyond the SM-like one can be canceled by the corresponding sparticle effects (and besides, are generally small because of the rather heavy $H,A$ and $H^\pm$ masses), the 2HDM has to be tested against a variety of such data. The     
$B$-physics observables that we have considered to that effect are listed  in Tab.~\ref{Tab:ExpResult}. We have computed the 2HDM predictions for these in all 2HDM Types using our own implementation, which output in fact agrees with the one from SuperIso \cite{Mahmoudi:2008tp}  (when run in 2HDM mode). 

\begin{table}[t!]
	\centering
	{\renewcommand{\arraystretch}{1.2}
		{\setlength{\tabcolsep}{0.15cm}
	\begin{tabular}{| c | c | c | c | c | c | c |}
		\hline
		$m_h$ (GeV)& $m_H$ (GeV)& $m_A$ (GeV)& $m_{H^\pm}$ (GeV)& $\alpha$ & 
		$\beta$& $m_{12}^2$ (GeV$^2$) \\
		\hline
		125  & [$m_{H^\pm}$; 1000] & [90; $m_{H^\pm}$] & [90; 1000] & [$\pi/5$; $\pi/2$] & [$-\pi/2$; $\pi/2$] & $m_A^2\tan\beta/(1+\tan^2\beta)$   \\
		\hline
	\end{tabular}}}
	\caption{Allowed range of variation for the free parameters of all 2HDM Types.}
	\label{table2}
\end{table}

\begin{figure}[h!]
	\includegraphics[width=8cm, height = 8cm]{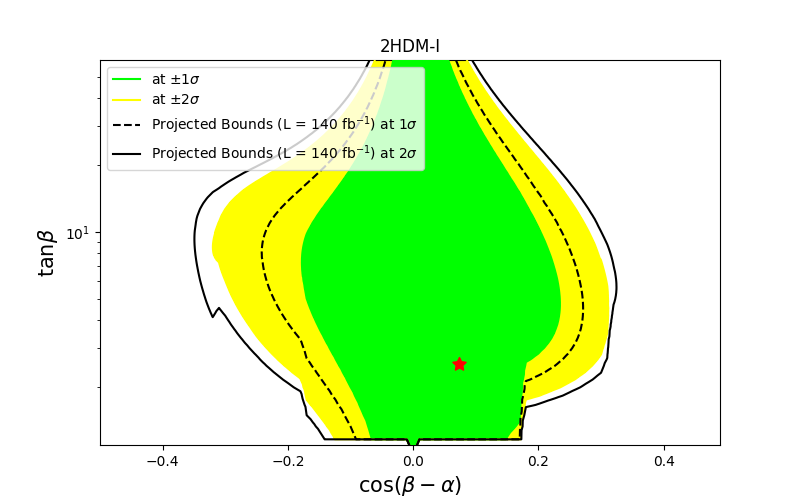}\includegraphics[width=8cm, height = 8cm]{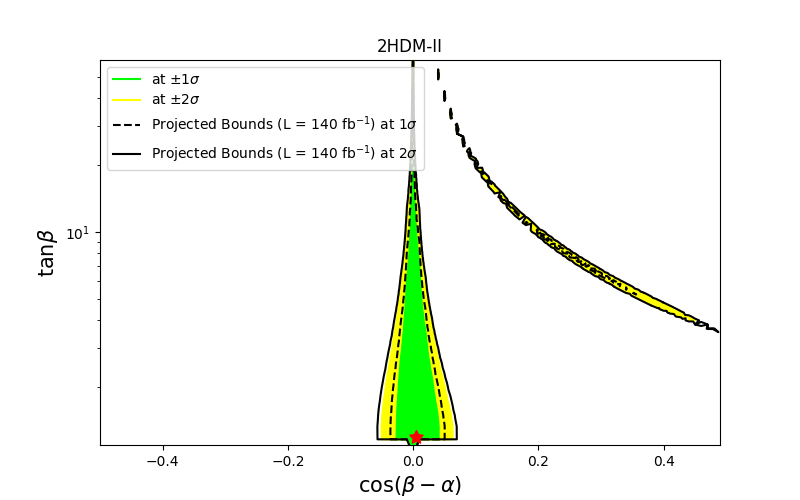}
	\includegraphics[width=8cm, height = 8cm]{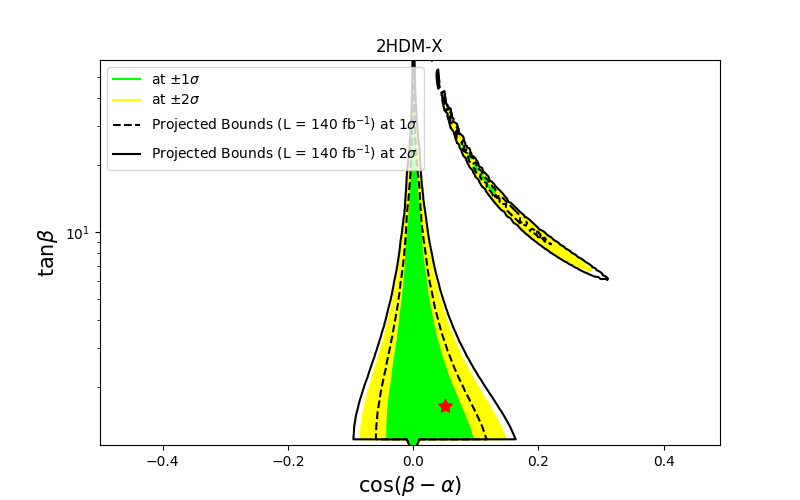}\includegraphics[width=8cm, height = 8cm]{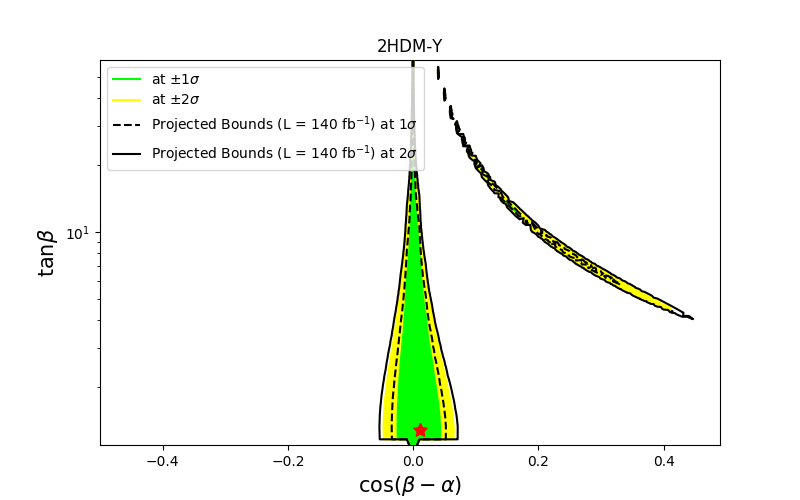}
	\vspace{0.2cm}
	\caption{Direct constraints from null heavy Higgs searches  at the LHC on the parameter space of the 2HDM Type-I (top left), Type-II (top right), Type-X (bottom left) and Type-Y (bottom right) mapped on the $(\alpha,\beta)$ plane. 
		The colors indicate compatibility with the observed Higgs signal at 1$\sigma$ (green), 2$\sigma$ (yellow) around the best fit points (red stars). }
	\label{2HDM-fig-1}
\end{figure}

Based on such constrained scans, we first illustrate in Fig.~\ref{2HDM-fig-1}, on the $(\cos(\beta-\alpha),\tan\beta)$ plane,  
the best fit points for the four 2HDM Types. Herein, are also shown the compatibility regions with the observed Higgs signal 
at the 1$\sigma$ (green) and 2$\sigma$  (yellow) level. The details of the best fit points herein (red stars) are given in 
Tab. \ref{table1} together with the values of the following observables: the total charged Higgs width $\Gamma_{H^\pm}$,
$\sigma(pp\to tH^-+{\rm c.c.})$, 
${\rm BR}(H^\pm\to \tau\nu)$,
${\rm BR}(H^\pm\to tb)$, ${\rm BR}(H^\pm\to A W^\pm)$ and ${\rm BR}(H^\pm\to hW^\pm)$. 
Note that in the 2HDM Type-II and -Y, the best fit point is located at a charged Higgs mass around 600 GeV
because of the $\overline{B}\to X_s \gamma$ constraints while in the 2HDM Type-I and- X one can fit data with a rather light charged Higgs state.

\begin{table}[h!]
	\begin{center}
		\begin{tabular}{|c|c|c|c|}
			\hline 
			\hline Observable 		&  Experimental result & SM contribution    & Combined at 1$\sigma$\\
			\hline ${\rm BR} (B\to\tau\nu)$									& $(1.14 \pm 0.22)\times 10^{-4}$~\cite{Amhis:2014hma}& $(0.78\pm 0.07 )\times 10^{-4}$ & $0.23\times 10^{-4}$\\
			\hline BR$(B^0_s \to \mu^+ \mu^-)$ 				& $(2.8\pm0.7)\times10^{-9}$~\cite{Archilli:2014cla} &$(3.66\pm0.28) \times 10^{-9}$ & $0.75\times 10^{-9}$ \\
			\hline BR$(B^0_d \to \mu^+ \mu^-)$& $(3.9\pm 1.5)\times 10^{-10}$~\cite{Archilli:2014cla}& $(1.08\pm 0.13) \times 10^{-10}$ & $1.50\times 10^{-10}$ \\
			\hline BR$(\overline{B}\to X_s\gamma)_{E_\gamma>1.6\,\text{GeV}}$ & $(3.43 \pm 0.22)\times 10^{-4}$~\cite{Amhis:2014hma}& $(3.36\pm 0.24) \times 10^{-4}$ &  $0.32\times 10^{-4}$\\
			\hline $\Delta M_s$ & $(17.757\pm 0.021)\text{ ps}^{-1}$~\cite{Agashe:2014kda,Amhis:2014hma}&$(18.257\pm 1.505) \text{ ps}^{-1}$ & $1.5\text{ ps}^{-1}$\\
			\hline $\Delta M_d$ & $(0.510\pm 0.003) \text{ ps}^{-1}$~\cite{Agashe:2014kda,Amhis:2014hma}	& $(0.548\pm0.075) \text{ ps}^{-1}$ & $0.075 \text{ ps}^{-1}$\\
			\hline 
			\hline
		\end{tabular}\end{center}
		\caption{Experimental results of flavor observables combined by the PDG and/or HFAG collaborations in Refs.~\cite{Amhis:2014hma,Agashe:2014kda}. As for BR$(B^0_q \to \mu^+ \mu^-)$, the combined results from the LHCb and CMS collaborations are shown as in Ref.~\cite{Archilli:2014cla}.}
		\label{Tab:ExpResult}
	\end{table}
	
	\begin{table}[t!]
		\centering
\hspace*{-1.0truecm}
		{\renewcommand{\arraystretch}{2} 
			{\setlength{\tabcolsep}{0.02cm}
				\begin{tabular}{|c | c | c | c | c |}
					\hline
					\hline
					Parameters  & Type-I & Type-II & Type-X & Type-Y \\
					\hline
					\hline
					($\alpha$, $\beta$)  & $(-0.30107, 1.19645)$  & $(-0.77474, 0.791554)$ & $(-0.49444, 1.02543)$  & $(-0.64861, 0.91044)$  \\
$(\cos(\beta-\alpha), \tan\beta)$ & $(0.07321, 2.54535)$  & $(0.00450, 1.01239)$ & $(0.05090, 1.64813)$  & $(0.01175, 1.28754)$ \\
\hline 
					($m_{H^\pm}$, $\Gamma_{H^\pm}$) (GeV) & (178, 1.4$\times10^{-2}$ ) & (592, 25.2) & (493, 7.63 ) & (631, 16.8) \\
					($m_A$, $m_H$) (GeV)& (97.71, 212) & (512, 694) & (412, 509) & (550, 652) \\
					${\rm BR}(H^\pm\to \tau\nu)$ & 0.4\%    &  -- & 0.03\%&  --\\
					${\rm BR}(H^\pm\to AW^\pm)$  &  55.2\%  &  0.05\% & 0.18\% &  0.08\% \\
					${\rm BR}(H^\pm\to hW^\pm)$ &  0.01\%  &   0.04\% & 0.9\% & 0.06\% \\
					${\rm BR}(H^\pm\to tb)$     & 44.1\%   &   99.7\%& 98.6\% & 99.6\% \\
					$\sigma(pp\to \bar tH^++{\rm c.c.})$ (fb)& 1570  & 434 & 308 & 214 \\
					\hline
					\hline
				\end{tabular}}}
				\caption{The best fit points in the 2HDM Type-I, -II, -X and -Y. The decay width $\Gamma_{H^\pm}$, cross sections $\sigma(pp\to tH^-+{\rm c.c.})$ 
					as well as relevant decay BRs for the  charged Higgs state are listed, for which values  smaller than 10$^{-4}$ are neglected. We have fixed $m_h=125$ GeV and $m_{12}^2=m^2_{A}\sin\beta \cos\beta$. 
				}
				\label{table1}
			\end{table}
			
			
			%

			\begin{figure}[h!]
				\includegraphics[width=8cm, height = 8cm]{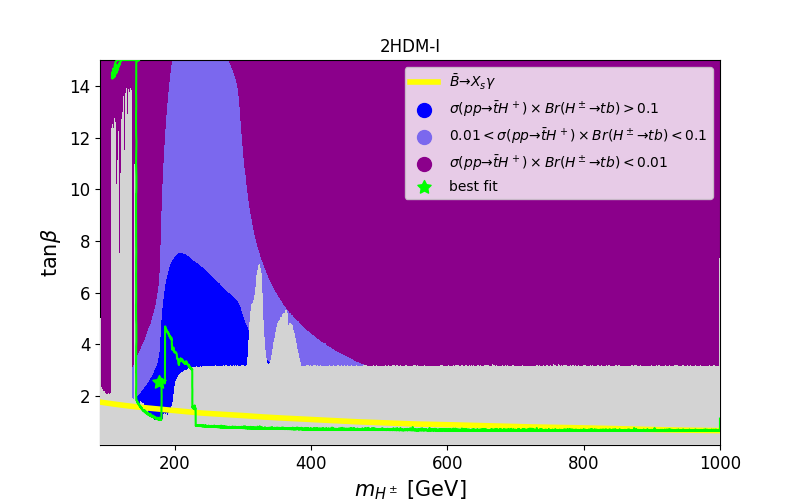}\includegraphics[width=8cm, height = 8cm]{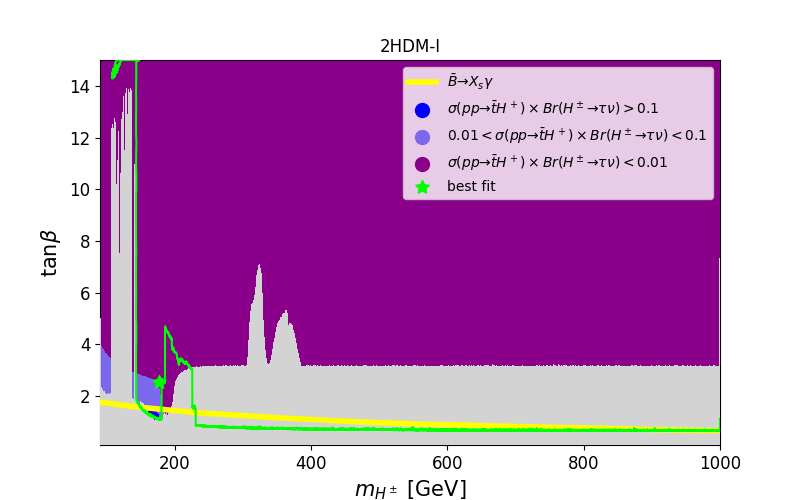}
				\includegraphics[width=8cm, height = 8cm]{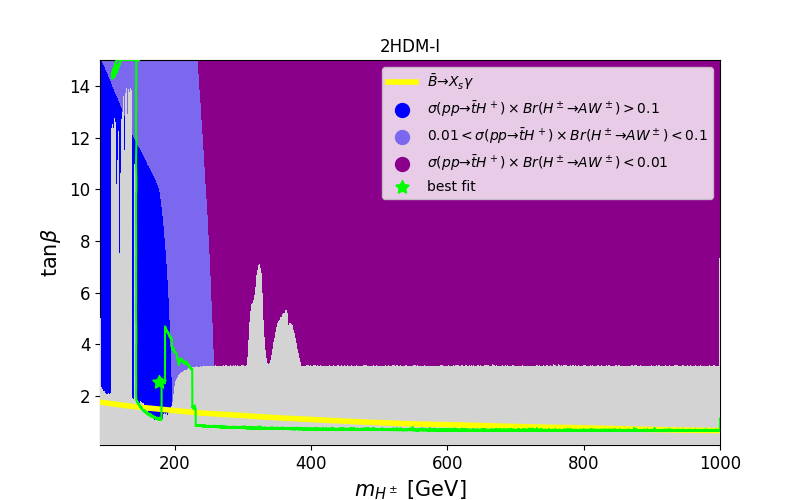}\includegraphics[width=8cm, height = 8cm]{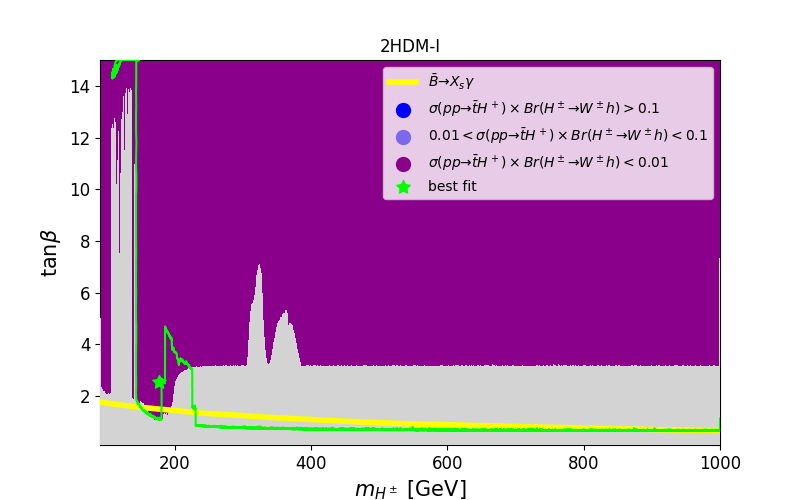}
				\vspace{0.2cm}
				\caption{The $\sigma(pp\to tH^-+{\rm c.c.})\times {\rm BR}(H^\pm\to XY)$ rate (in pb) at $\sqrt{s}$=14 TeV in the 2HDM Type-I, for $XY\equiv tb$ (top left), $XY\equiv \tau\nu$ (top right), $XY\equiv AW^\pm$ (bottom left) and $XY\equiv hW^\pm$ (bottom right). Exclusion bounds at 95\% CL from the non-observation of the additional Higgs states are overlaid in gray. The green contour indicates compatibility with the observed Higgs signal at $68\%$ CL and the best fit (benchmark) points are marked by green stars. The solid yellow line contours are the boundary of 
					95\% CL exclusion from $\overline{B}\to X_s\gamma$ measurements. The maximum of the cross section $\sigma(pp\to tH^-+{\rm c.c.})\times {\rm BR}(H^\pm\to XY)$ is 3.1 pb and 1.83 pb for $XY\equiv AW^\pm$, $tb$, respectively.}
				\label{figure:type1}
			\end{figure}
			
\begin{figure}[h!]
				\includegraphics[width=8cm, height = 8cm]{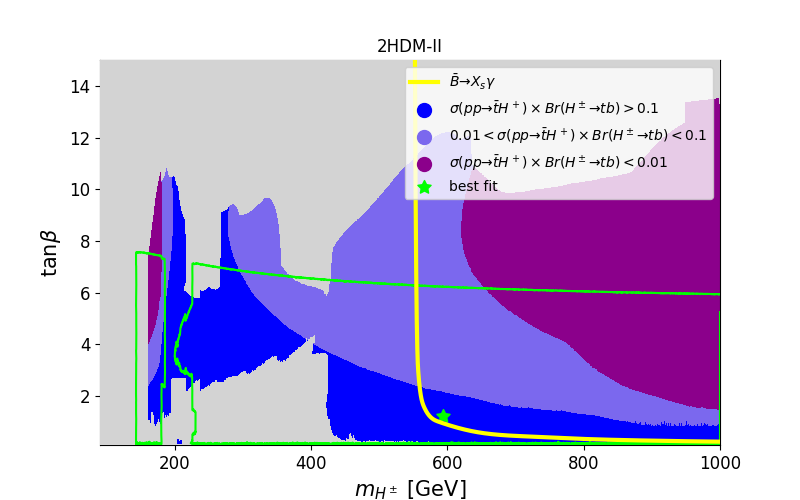}\includegraphics[width=8cm, height = 8cm]{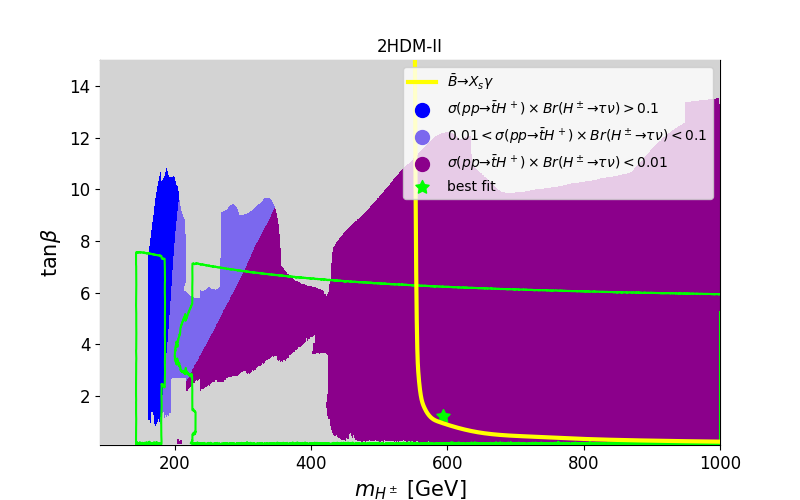}
				\includegraphics[width=8cm, height = 8cm]{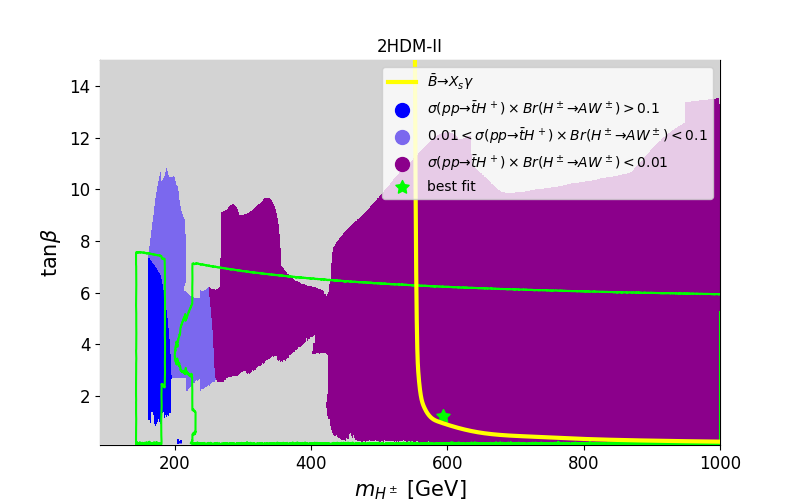}\includegraphics[width=8cm, height = 8cm]{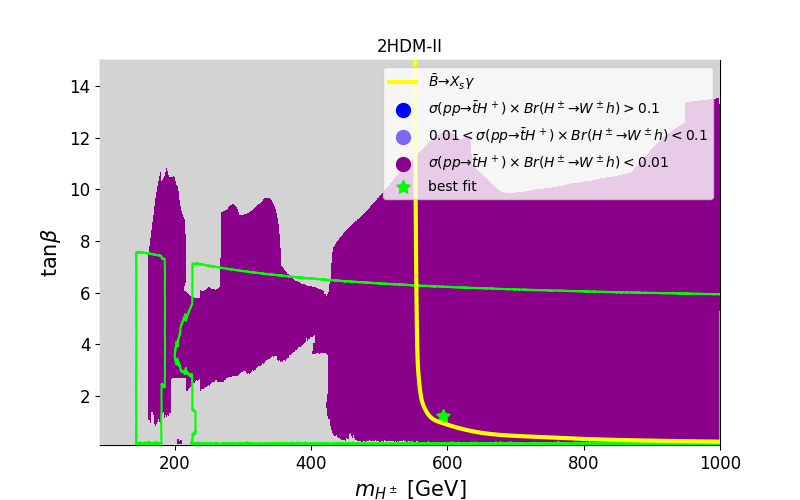}
				\vspace{0.2cm}
				\caption{The $\sigma(pp\to tH^-+{\rm c.c.})\times {\rm BR}(H^\pm\to XY)$ rate (in pb) at $\sqrt{s}$=14 TeV in the 2HDM Type-II, for $XY\equiv tb$ (top left), $XY\equiv \tau\nu$ (top right), $XY\equiv AW^\pm$ (bottom left) and $XY\equiv hW^\pm$ (bottom right). The maximum of the cross section $\sigma(pp\to tH^-+{\rm c.c.})\times {\rm BR}(H^\pm\to XY)$ is 2.3 pb and 1.38 pb for $XY\equiv AW^\pm$, $tb$, respectively. The color  coding is the same as in Fig.~\ref{figure:type1}.}
				\label{figure:type2}
			\end{figure}

\begin{figure}[h!]
				\includegraphics[width=8cm, height = 8cm]{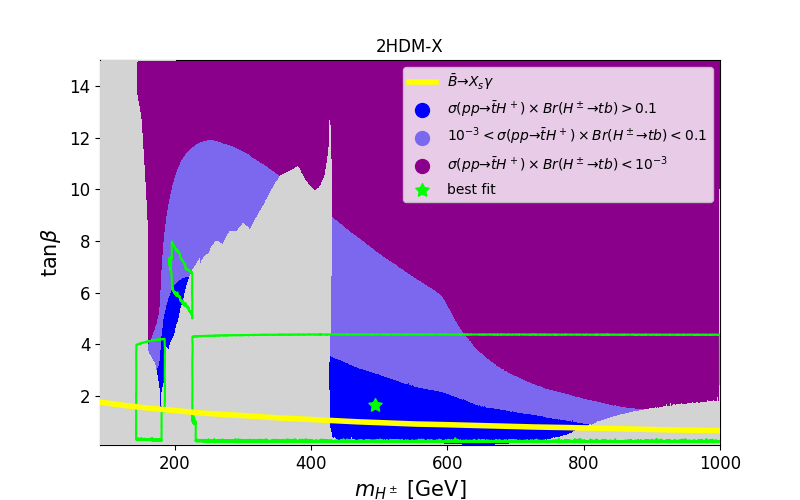}\includegraphics[width=8cm, height = 8cm]{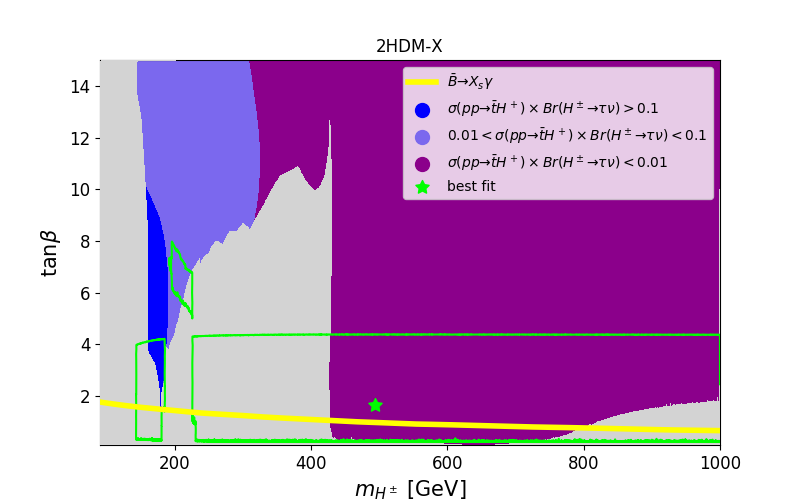}
				\includegraphics[width=8cm, height = 8cm]{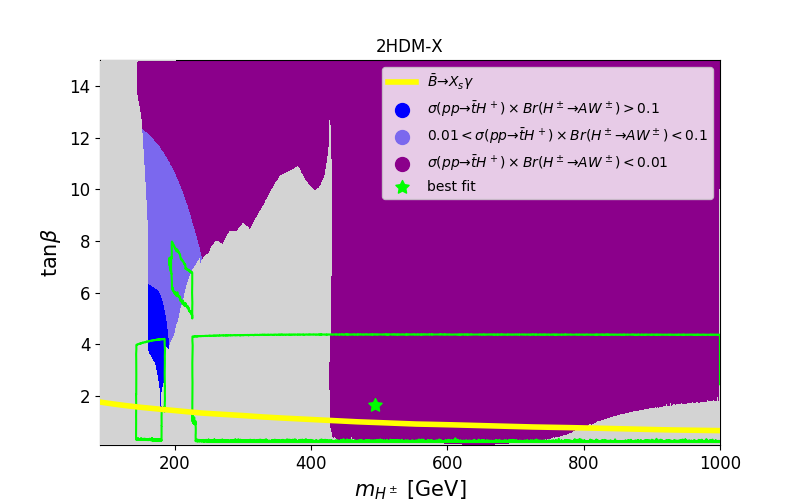}\includegraphics[width=8cm, height = 8cm]{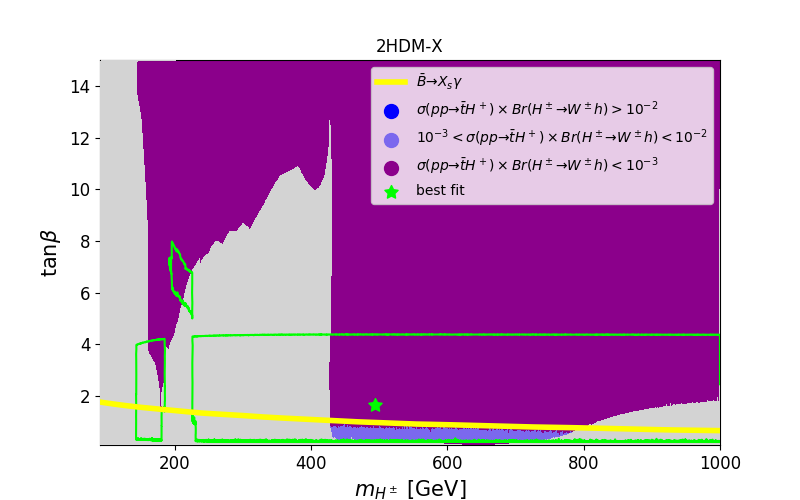}
				\vspace{0.2cm}
				\caption{The $\sigma(pp\to tH^-+{\rm c.c.})\times {\rm BR}(H^\pm\to XY)$ rate (in pb) at $\sqrt{s}$=14 TeV in the 2HDM Type-X, for $XY\equiv tb$ (top left), $XY\equiv \tau\nu$ (top right), $XY\equiv AW^\pm$ (bottom left) and $XY\equiv hW^\pm$ (bottom right). The maximum of the cross section $\sigma(pp\to tH^-+{\rm c.c.})\times {\rm BR}(H^\pm\to XY)$ is 2.3 pb and 1.23 pb for $XY\equiv AW^\pm$, $tb$, respectively. The color  coding is the same as in Fig.~\ref{figure:type1}.}
				\label{figure:type3}
			\end{figure}

\begin{figure}[h!]
				\includegraphics[width=8cm, height = 8cm]{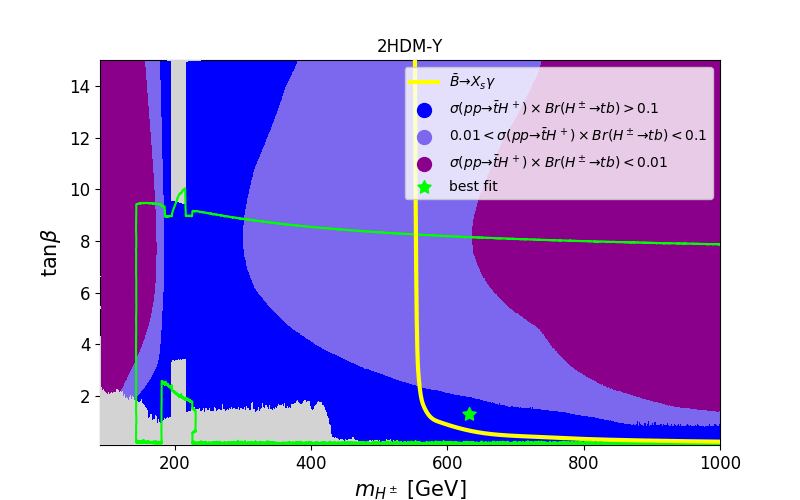}\includegraphics[width=8cm, height = 8cm]{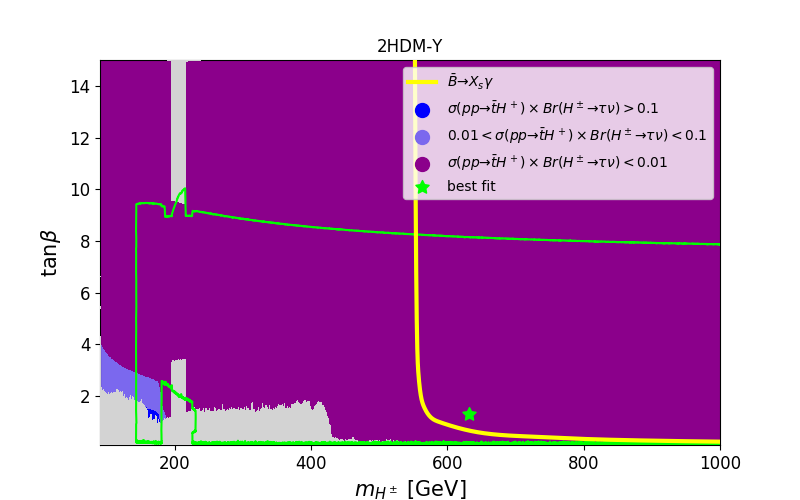}
				\includegraphics[width=8cm, height = 8cm]{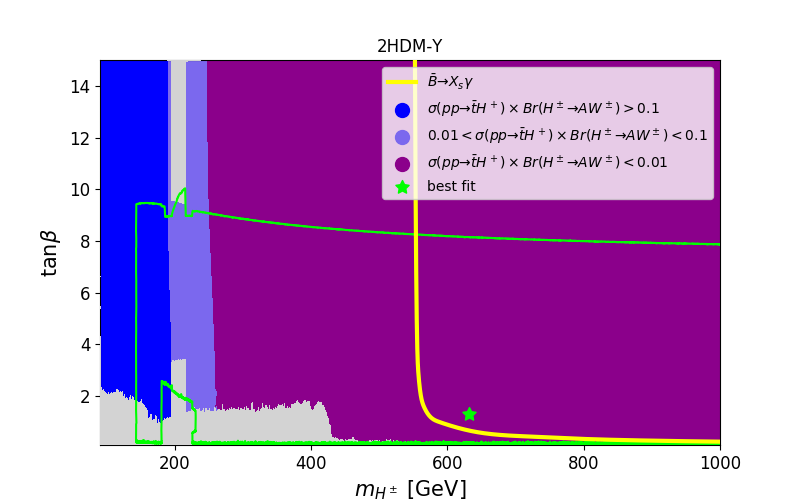}\includegraphics[width=8cm, height = 8cm]{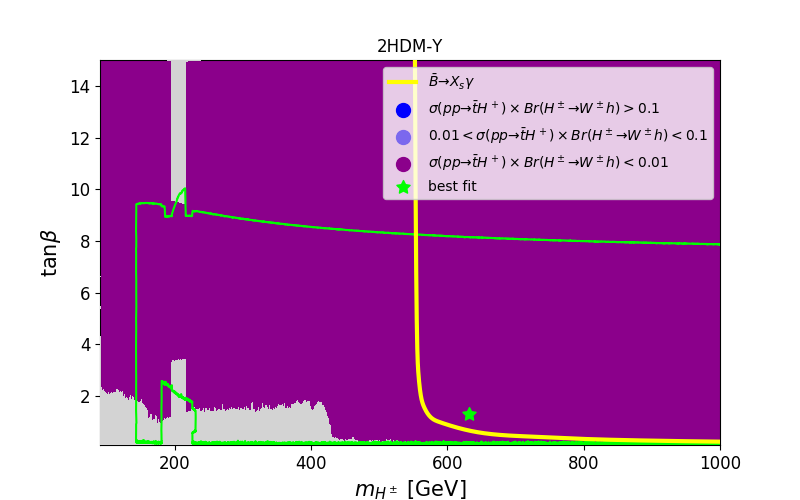}
				\vspace{0.2cm}
				\caption{The $\sigma(pp\to tH^-+{\rm c.c.})\times {\rm BR}(H^\pm\to XY)$ rate (in pb) at $\sqrt{s}$=14 TeV in the 2HDM Type-Y, for $XY\equiv tb$ (top left), $XY\equiv \tau\nu$ (top right), $XY\equiv AW^\pm$ (bottom left) and $XY\equiv hW^\pm$ (bottom right). The maximum of the cross section $\sigma(pp\to tH^-+{\rm c.c.})\times {\rm BR}(H^\pm\to XY)$ is 3.54 pb and 1.85 pb for $XY\equiv AW^\pm$, $tb$, respectively. The color  coding is the same as in Fig.~\ref{figure:type1}.}
				\label{figure:type4}
			\end{figure}

In Fig.~\ref{figure:type1}(Fig.~\ref{figure:type2})[Fig.~\ref{figure:type3}]\{Fig.~\ref{figure:type4}\},  
we show (in gray) over the  ($m_{H^\pm}$, $\tan\beta$) plane the  95\% CL exclusion region from the 
non-observation of the additional Higgs states for 2HDM Type-I(-II)[-X]\{-Y\}. In all these plots, we also draw 
(as a solid yellow line) the 95\% CL exclusion from ${\rm BR}(\overline{B} \to X_s\gamma)$ together with 
a solid green line  representing the 1$\sigma$ compatibility with the Higgs signals observed at the LHC. 
As a green star, we also give the best fit point to these data over the available parameter space for all Types 
(these are the same as the red stars in the previous figure).
It is clear from these plots that, in the 2HDM-I and -X, one can still have relatively light charged Higgs states
 (of the order 100 to 200 GeV in mass) that are consistent with all aforementioned data, crucially including 
$B$-physics observables.  In addition, such light charged Higgs does not affect too much the rate of $h\to \gamma\gamma$
which is strongly dominated by the $W^\pm$ loops while the charged Higgs loops are subleading.
In the case of the 2HDM Type-II and -Y, the BR$(\overline{B}\to X_s\gamma$) constraint  pushes the charged Higgs  boson 
mass to be higher than 580 GeV. (Note that, in the 2HDM Type-II, it is clear that, like for the MSSM case, large $\tan\beta$ is 
excluded mainly from $H,A\to \tau^+ \tau^-$ as well as from $H^+\to \tau\nu$ searches at the LHC).
However, for 2HDM Type-X, one can see  that light charged Higgs states, with $m_{H^\pm}$ $\leq 170$ GeV, 
are excluded for all $\tan\beta$'s and this is due to charged Higgs searches failing to detect $H^\pm\to$ $\tau\nu$.
			
			We now discuss the size of the charged Higgs production cross section times its BRs in decay channels such as $H^+\to  \tau \nu, t\bar{b}, AW^\pm$ and  $hW^\pm$.
			In Fig.~\ref{figure:type1}(top-left panel) we illustrate  the values of 
			$\sigma(pp\to tH^-+{\rm c.c.})\times {\rm BR}(H^\pm\to tb)$ (in pb) where we can see that it is possible to 
			have a production times decay rate in the range $0.01$ to $0.2$ pb  for $1\leq \tan\beta\leq $ {6} and $180$ GeV $< m_{H^\pm}< 300$ GeV.
			This could lead to more than thousands raw  $t\bar{t}b$ signal events for 100 fb$^{-1}$ luminosity. 
			In the case of $H^\pm \to \tau \nu$ and $H^\pm \to h W^\pm $, which are suppressed, respectively,  by $1/\tan\beta$
			and $\cos(\beta-\alpha)\approx 0$, the rate is much smaller than for the $tb$ mode.
			In contrast, since the coupling $H^\pm W^\mp A$ is a gauge coupling without any 
			suppression factor, when  $H^\pm \to AW^\pm$ is open, it may dominate over the $H^\pm \to tb$ channel.
			One can see from Fig.~\ref{figure:type1}(bottom-left panel) that, for $100$ GeV  $< m_{H^\pm}< 220$ GeV
			and for  $1\leq \tan\beta\leq 14$, the corresponding rate for 
			$\sigma(pp\to tH^-+{\rm c.c.})\times {\rm BR}(H^\pm\to AW^\pm) \geq 0.01$ pb. This could lead to an interesting final state 
			$b W^+W^-A$ where one $W^\pm$ could decay leptonically, hence offering a clean trigger. The decay $H^\pm\to hW^\pm$ is essentially inaccessible, see Fig.~\ref{figure:type1}(bottom-right panel).
			
			In the case of 2HDM Type-II and -Y, as one can see from Fig.~\ref{figure:type2} and 
			Fig.~\ref{figure:type4},  respectively,  there is a wide region over the ($m_{H^\pm}$, $\tan\beta$) plane where 
			the rate for $\sigma(pp\to tH^-+{\rm c.c.})\times {\rm BR}(H^\pm\to tb)$ is rather sizable for both  
			moderate ($m_{H^\pm }\leq 300$ GeV) and heavy (otherwise) charged Higgs masses (top-left panel). However, if one takes into account the
			$\overline{B}\to X_s\gamma$ constraint, then $m_{H^\pm}$ is required to be much heavier than  580 GeV (as already discussed), 
			which makes the rate $\sigma(pp\to tH^-+{\rm c.c.})\times {\rm BR}(H^\pm\to tb)\geq 0.1$ pb
			only for $\tan\beta < 1.5$. All the other channels (in the three remaining panels) have smaller production times decay rates.
			
			The 2HDM Type-X is depicted in Fig.~\ref{figure:type3}, wherein the usual production times BR rates are shown. The top-right panel is again for the $H^+\to t\bar{b}$ channel, which {exhibits} a potentially interesting cross section ($\geq 1$ fb)  in the $H^+ \to t\bar{b}$ channel for both a light charged Higgs mass 
			(around 200 GeV) and a heavy one (around $420$ GeV). In the case of the $\tau \nu$ channel (top-right panel),
			one can get {sizable} rates for $\sigma(pp\to tH^-+{\rm c.c.})\times {\rm BR}(H^\pm\to \tau \nu)$ for a charged Higgs mass around 
			200 GeV and $\tan\beta\geq  2$.

In all 2HDM Types, we elect the best fit points to also be the BPs amenable to experimental tests by ATLAS and CMS.

\section{Conclusions}
We have studied charged Higgs boson phenomenology  in both the MSSM and 2HDM, the purpose being to define BPs amenable to phenomenological  investigation already with the full Run-1 and -2 datasets and certainly accessible with the Run-3 one of the LHC.  They have been singled out following the enforcement of the latest theoretical and experimental constraints, so as to be entirely up-to-date. Furthermore, they have been defined with the intent of increasing sensitivity of dedicated (model-dependent)  $H^\pm$ searches 
to some of the most probable parameter space configurations of either scenario. With this in mind, we have listed in two tables their input and output values, the former in terms of the fundamental parameters of the model concerned and the latter in terms of key observables (like, e.g., physical masses and couplings, production cross sections and decay BRs). We have also specified which numerical tools we have used to produce all such an information, including their settings. 

For the MSSM we have concentrated on two popular scenarios, i.e., the  hMSSM and $m_{h}^{\rm mod+}$ ones.
It was found that the hMSSM case still possesses a rather large available parameter space, here mapped over the $(m_A,\tan\beta)$ plane,  
while the $m_{h}^{\rm mod+}$ one is instead much more constrained. In terms of the largest production and decay rates,  
in the hMSSM scenario  one finds that the most copious channels, assuming $pp\to tH^-$ + c.c. production, are via the decay $H^+\to t\bar{b}$ followed by $H^+\to \tau \nu$ whereas  for the $m_{h}^{\rm mod+}$ scenario the decay modes $H^+\to t\bar{b}$ and $H^+\to \tilde{\chi}_1^+ \tilde{\chi}_0$ 
offer the largest rates. In both cases, only $m_{H^\pm}>m_t$ values are truly admissible by current data.

Within the 2HDM, we have looked at at the four standard Yukawa setups, known as Type-I, -II, -X and -Y. Because of $\overline{B}\to X_s \gamma$ constraints, the profile of a charged Higgs in the 2HDM Type-II and -Y is a rather heavy one, with a mass required to be more than 580 GeV.  While this puts an obvious limit to LHC sensitivity owing to a large phase space suppression in production, we have emphasized that $H^\pm\to b\bar b W^\pm$ channels should be searched for, with intermediate contributions from the $AW^\pm$ and $tb$ modes (including their interference \cite{Arhrib:2017veb}), alongside $H^\pm\to \tau\nu$.  In the case of the 2HDM Type-I and -X, a much lighter charged Higgs state is still allowed by data, in fact, even with a mass below that of the top quark. While the configuration $m_{H^\pm}<m_t$ is best probed by using $t\bar t$ production and decays into $\tau\nu$, the complementary mass region, i.e., $m_{H^\pm}>m_t$ (wherein $pp\to tH^-$ + c.c. is the production mode), may well be accessible via a combination of $H^+\to t\bar{b}$ and 
$H^\pm\to AW^\pm$ (in Type-I) plus $H^\pm\to \tau\nu$ (in Type-X). 

\section*{Note added}

{{Since the original submission of this paper, several new
    experimental analyses have been carried out by ATLAS and CMS using the
    full Run-2 data sample of $\approx139$ fb$^{-1}$. Some of these, covering
    both measurements of the SM-like Higgs boson and the search for new
    (pseudo)scalar Higgs states, both charged and neutral, have been captured
    by the latest versions of HiggsBounds and HiggsSignals, 
HiggsBounds-5.3.2beta  and  HiggsSignals-2.2.3beta , 
respectively. Likewise, further analyses by LHCb of flavor observables have been carried out since and most of these have been captured by the latest version of SuperIso. Hence, we have repeated our scans using all such tools and found negligible differences between our original results and the new ones. Further, we have investigated which ones of the full Run-2 data set analyses were not incorporated in the above codes and found that their ad-hoc application to our analysis  did not change our results either.}

\section*{Acknowledgments}
AA, RB and SM are supported by the grant H2020-MSCA-RISE-2014 no.  645722 (NonMinimalHiggs).  This work is
also supported by the Moroccan Ministry of Higher Education and Scientific Research MESRSFC and CNRST: Project
PPR/2015/6.  SM is supported in part through the NExT. Institute and the STFC CG ST/L000296/1. 
For the avoidance of doubt, we acknowledge that this paper has already been submitted to a public database as https://arxiv.org/abs/1810.09106.

\bibliographystyle{frontiersinHLTH&FPHY} 

\end{document}